\documentclass[prl,aps,superscriptaddress,notitlepage,longbibliography,twocolumn,footinbib,superscriptaddress]{revtex4-2}
\usepackage{amsmath}
\usepackage{amsfonts}
\usepackage{graphicx}
\usepackage{bm}
\usepackage{bbold}
\usepackage{color}
\usepackage{soul}
\usepackage{braket}
\usepackage{epstopdf}
\long\def\comment#1{}

\usepackage[dvipsnames]{xcolor}

\newcommand\red{}

\usepackage[colorlinks]{hyperref}

\hypersetup{
    colorlinks=true,
    linkcolor=blue,
    filecolor=magenta,      
    urlcolor=magenta,
    citecolor={blue},
    }

\DeclareMathAlphabet{\mathcallc}{U}{dutchcal}{m}{n}
\SetMathAlphabet{\mathcallc}{bold}{U}{dutchcal}{b}{n}
\DeclareMathAlphabet{\mathbcallc}{U}{dutchcal}{b}{n}

\vspace*{-\baselineskip}

\begin{document}

\title{What is the spectral density of the reservoir for a lossy quantized cavity?}
\author{Chris Gustin}
\affiliation{E.\,L.\,Ginzton Laboratory, Stanford University, Stanford, California 94305, USA}
\email{cgustin@stanford.edu}
\author{Juanjuan Ren}
\affiliation{\hspace{0pt}Department of Physics, Engineering Physics, and Astronomy, Queen's University, Kingston, Ontario K7L 3N6, Canada\hspace{0pt}}
\author{Stephen Hughes}
\affiliation{\hspace{0pt}Department of Physics, Engineering Physics, and Astronomy, Queen's University, Kingston, Ontario K7L 3N6, Canada\hspace{0pt}}

\begin{abstract}
 By considering a two-level system weakly coupled to a single lossy three-dimensional cavity mode,
 and using a quasinormal mode expansion
 for the photonic medium, we show that the 
 frequency-dependence of the quantum coupling interaction between the cavity and its photonic reservoir can not be simply constructed to be independent 
 of the cavity contents. For the single 
 quantum dipole case, we identify the correct form of the \red{local} spectral density, revealing a surprising \red{gauge-dependent} \red{$\sim \omega^{\pm 1}$} prefactor scaling, as well as a spatially-dependent correction.
We thus establish the correct quantum form for the cavity-reservoir interaction, which so far has only been worked out rigorously for simple 1D geometries, and show its significant impact on broadband strong coupling.

\end{abstract}

\maketitle

\emph{Introduction.}
One of the most fundamental concepts in quantum optics is 
that of the quantized cavity. By spatially confining electromagnetic (EM) fields, the continuous nature of the fields' degrees of freedom can be collapsed into discrete resonances, modelled as harmonic oscillator modes. 
The simplicity of this model and ease of physical implementation has made quantum optics an indispensable testbed for many 
aspects of quantum mechanics, 
from fundamental science~\cite{Giustina2015Dec,Shalm2015Dec} 
to 
quantum information applications~\cite{Mabuchi2002Nov,Weedbrook2012May,Browne2017,Lambropoulos2000,RevModPhys.87.1379,RevModPhys.87.347,Haroche2020}.

Despite the ubiquity of this quantized cavity model, the \emph{ab initio} fundamental description of the underlying  reality is 
subtle and 
complex, and the complete details of how such a collapse into discrete quantized modes can be recovered from an underlying quantization of the EM fields in media remains an outstanding challenge. This is because replacing continua with discrete resonant modes relies on an assumption of a \emph{closed system}, where photons remain localized in the cavity modes indefinitely. However, in reality, all cavities are leaky cavities.

A major advance in modelling this reality came with 
the \emph{input-output} and \emph{system-reservoir} theory~\cite{Yurke1984Mar,Gardiner1985Jun}, where 
the discrete cavity modes couple via a Hamiltonian term to a continuous ``reservoir'' of photon modes. Under reasonable assumptions, a Lindblad master equation can be derived for the subspace of discrete modes, giving Markovian photon loss. Moreover, the quantum statistics of the photons inside the cavity can be directly related to those which are lost from the cavity by means of a time-local ``input-output'' relationship.

Although such models have since found 
widespread application in quantum optics, they still rely on phenomenological assumptions on the form of the system-reservoir interaction. For example, specifying their general analysis to the case of a lossy single-mode cavity, 
standard input-output 
theories 
posit the form of the interaction Hamiltonian $\hat{H}_{\rm int}$ as (using $\hbar=1$ throughout)
\begin{equation}\label{eq:hint}
    \hat{H}_{\rm int} = \int_{-\infty}^{\infty} d\omega \Lambda(\omega)\left[\hat{a}\hat{b}^{\dagger}_{\omega} + \hat{a}^{\dagger}\hat{b}_{\omega}\right],
\end{equation}
where $[\hat{b}_{\omega},\hat{b}^{\dagger}_{\omega'}] = \delta(\omega-\omega')$, and $[\hat{a},\hat{a}^{\dagger}]=1$. The functional form of the \emph{system-bath coupling function} $\Lambda(\omega)$
(related to the spectral density $J(\omega)= 2\pi |\Lambda(\omega)|^2$) is left unspecified, as the only value that enters into the final calculations, after a Markov approximation, is $|\Lambda(\omega_c)|$, with $\omega_c$  the resonant cavity frequency. Typically, one assumes that the form of $\Lambda(\omega)$ is thus not important, at least in instances where the rotating-wave approximation holds [i.e., the ``empty'' cavity of Fig.~\ref{fig:schematic}(a)]. Introducing a strong dipole-coupling of a two-level system (TLS) to the cavity can allow for a surpassing of this regime towards the ultrastrong-coupling (USC) regime of quantum electrodynamics~\cite{frisk_kockum_ultrastrong_2019,forn-diaz_ultrastrong_2019}, where  
it is now well-established that the form of $\Lambda(\omega)$ 
{\it does} matter a great deal when it comes to predicting observable outputs of the cavity-TLS system, including emission spectra~\cite{Salmon2022Mar,Hughes2023Sep,Bamba2014Feb} [Fig.~\ref{fig:schematic}(b)]. 

To go beyond a phenomenological approach as in Eq.~\eqref{eq:hint} to general ``few-mode'' scattering problems \red{(that is, to construct \emph{ab initio} a theory of system-reservoir coupling where the total Hamiltonian is derived, rather than phenomenologically posited)} has also been highlighted in other contexts recently~\cite{Lentrodt2020Jan,Lentrodt2020Jun}. Progress has been made in this manner in certain 1-D systems~\cite{Dutra2000Oct,PhysRevA.88.043819,PhysRevA.72.053813}, as well as dispersionless systems~\cite{Viviescas2003Jan}.
The development of a theory consistent with causality (including material dispersion and absorption) which is valid and tractable for general, 3-D resonators, however, remains an outstanding challenge. Pseudomode approaches offer a potential solution, but often involve the introduction of many additional modes beyond the actually present resonances in the system~\cite{Lednev2023May,Lambert2019Aug}.

In this work, we \red{show}
that such a frequency dependence of the system-bath coupling \red{can produce} observable effects \red{outside of} the USC regime, and in doing so, make a major advance towards a more rigorous theory of quantized cavity loss for general systems. We show that a resolution to the question of the system-bath coupling interaction form is intrinsically connected to the complex-valued open {\it quasinormal} mode (QNM) of the cavity. Our 
general findings are: (i) for cavity-QED systems with a single quantized mode interacting with a TLS dipole, if the mode can be described by a 
purely 
real-valued QNM at the dipole location, the \red{local} system-bath coupling scales as $\Lambda(\omega) \sim \omega^{\red{\pm \frac{1}{2}}}$\red{, depending on the gauge.} (ii) For \emph{realistic} cavity modes, the product of the QNM $Q$-factor and phase ($\phi$) at any location  is typically non-negligible for the purposes of calculating corrections beyond phenomenological loss models, which leads to a \emph{failure} of 
traditional system-bath coupling models to capture the correct broadband dynamics, for which we present a heuristic fix.

We consider a weakly-coupled system of a single TLS and cavity mode interacting within the dipole approximation. By employing the rigorous theory of gauge-invariant macroscopic QED~\cite{Dung1998May,Gustin2023Jan}, we utilize a QNM expansion of the medium (cavity + environment) Green's function to determine the frequency dependence of a TLS subject to radiative decay in a frequency window dominated by a single mode, which we also verify with a full numerically exact calculation of the Green's function.

\begin{figure}[htb]\label{fig:schematic}
    \centering
    \includegraphics[width = 0.99\columnwidth]{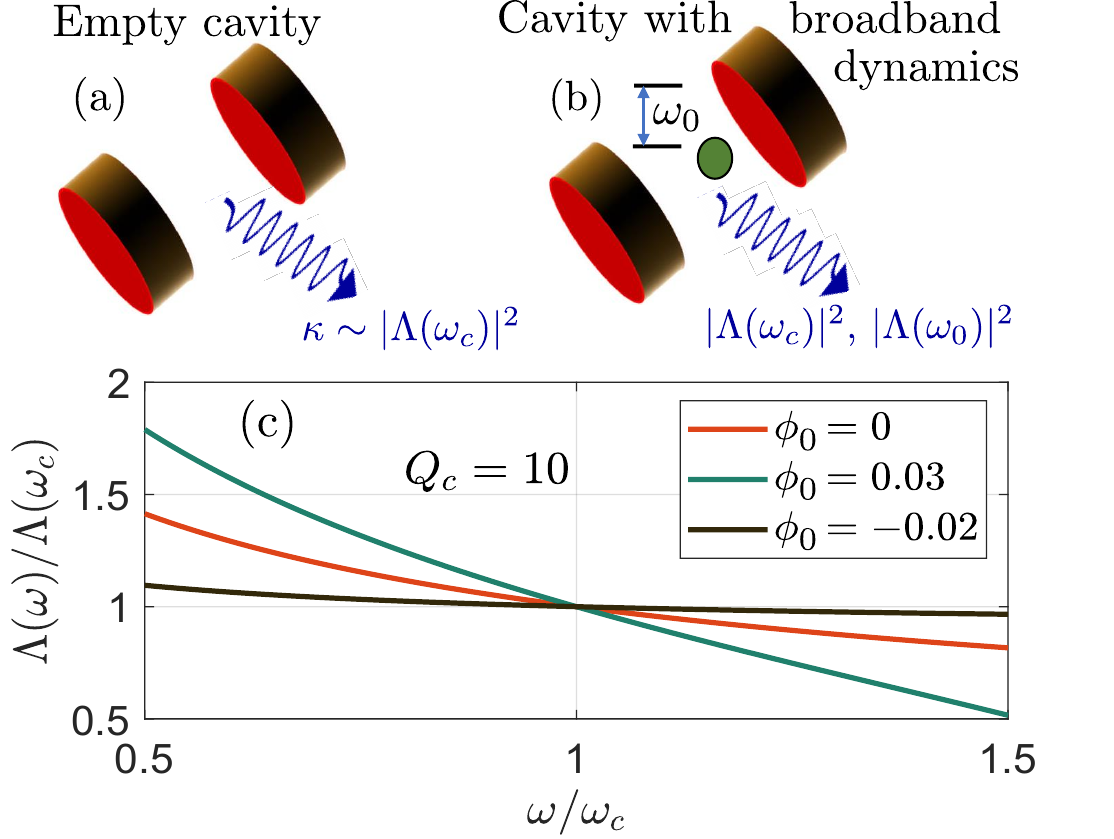}      
    \caption{Illustration of the differences between loss in (a) an empty cavity and (b) one undergoing broadband dynamics. When internal dynamics occur only at the resonant frequency, the only quantity relevant in dissipation is 
    one 
    parameter $|\Lambda(\omega_c)|$. In contrast, under broadband dynamics (here illustrated by a detuned and weakly-coupled TLS with frequency $\omega_0$), the spectral density is probed for different frequency values, highlighting the importance of its functional dependence. Shown in (c) is the spectral density identified later in Eq.~\eqref{eq:heuristic} for various values of the QNM phase $\phi_0$ at the TLS location. 
    }
    \label{fig:schematic}
\end{figure}

Comparison with the quantized cavity-photon reservoir model yields the correct choice of system-bath coupling $\Lambda(\omega)$.
Interestingly, {\it this result is gauge-dependent}. \red{Specifically, the correct form of $\Lambda(\omega)$ depends on the gauge in which the system-bath coupling is assumed to be written as cavity operators coupled to a bath.}
\red{These results indicate} that gauge-dependent predictions in cavity quantum electrodynamics---which, in the case of gauge non-invariance caused by truncation, the resolution to which is now well known~\cite{DiStefano2019Aug,Taylor2020Sep,Settineri2021Apr,Savasta2021May,Salmon2022Mar,Taylor2022May,Gustin2023Jan}\red{---}are not only perceptible in extreme regimes of light-matter coupling (USC), but are also relevant whenever a broadband frequency dependence is present.

\emph{Continuum Theory.}
As a representative system, we consider a dipole-TLS weakly coupled to a single resonant cavity mode, and derive the spontaneous emission (SE) rate of the dipole. 
First, we
consider the photonic medium (cavity) to consist of a quantized continuum of reservoir operators, corresponding to the (complex) dielectric medium of the cavity and environment. To this end, we employ the macroscopic QED formalism, which is appropriate for arbitrary dispersive and absorbing media. 
The transverse part of the (manifestly gauge-invariant) vector potential of the EM fields is~\cite{Gustin2023Jan}
\begin{align}
    \hat{\mathbf{A}}_{\perp}(\mathbf{r}) = \!\int  \! \! d^3r' & \! \! \int_0^{\infty} \! \! \frac{d\omega}{\omega}\sqrt{\frac{\epsilon_I(\mathbf{r'},\omega)}{\epsilon_0 \pi}} \nonumber \\ & \times \mathbf{G}_{\perp}(\mathbf{r},\mathbf{r'},\omega) \cdot \hat{\mathbf{b}}(\mathbf{r'},\omega) + \text{H.c.},
\end{align}
where $\epsilon_I$ is the imaginary part of the medium's dielectric function, and $\mathbf{G}_{\perp}(\mathbf{r},\mathbf{r'},\omega)$ is the medium's Green's function, transverse with respect to the left-hand side of the dyad, and spatial argument $\mathbf{r}$. The operators that correspond to the EM fields' polariton excitations 
satisfy
$[\hat{b}_i(\mathbf{r},\omega),\hat{b}^{\dagger}_j(\mathbf{r'},\omega')] = \delta_{ij}\delta(\mathbf{r}-\mathbf{r'})\delta(\omega-\omega')$.

In the Coulomb gauge~\cite{Gustin2023Jan}, the Hamiltonian for the TLS interacting with the \red{transverse} EM fields is
\begin{equation}\label{eq:HC}
\hat{H} = \hat{H}_{\rm F} + e^{i \mathbf{d} \cdot \hat{\mathbf{A}}_{\perp,0}\hat{\sigma}_x}
\hat{H}_0  e^{-i \mathbf{d} \cdot \hat{\mathbf{A}}_{\perp,0}\hat{\sigma}_x},
\end{equation}
with $\hat{\mathbf{A}}_{\perp,0} = \hat{\mathbf{A}}_{\perp}(\mathbf{r}_0)$, where $\mathbf{r}_0$ is the dipole location, $\hat{H}_{\rm F} = \int d\omega \int d^3r \omega \hat{\mathbf{b}}^{\dagger}(\mathbf{r},\omega) \cdot \hat{\mathbf{b}}(\mathbf{r},\omega)$, $\hat{H}_0 = \omega_0 \hat{\sigma}^+\hat{\sigma}^-$, and $\mathbf{d}$ is the (assumed real) dipole moment. The Pauli operators $\hat{\sigma}^{\pm}$ correspond to raising/lowering operators of the TLS~\footnote{\red{Here we have neglected longitudinal field couplings~\cite{Gustin2023Jan,Knoll2000Jun} for simplicity, assuming the transverse couplings to dominate for the single-mode systems of interest, which we also verify in the Supplementary Material~\cite{SI}}}.
Assuming non-ultrastrong coupling, we expand Eq.~\eqref{eq:HC} to first order in $\mathbf{d} \cdot \hat{\mathbf{A}}_{\perp,0}$. This gives
\begin{equation}
    \hat{H} = \hat{H}_{\rm F} + \hat{H}_0 + \omega_0 \mathbf{d} \cdot \hat{\mathbf{A}}_{\perp,0}\hat{\sigma}_y.
\end{equation}
One can then derive the master equation of the TLS subsystem, using standard second-order Born-Markov techniques~\cite{Breuer2007Mar}. The result is (neglecting Lamb shifts and making a rotating-wave approximation) a Lindblad contribution to the master equation $\gamma\left[\hat{\sigma}^{-}\hat{\rho}_{\rm S} \hat{\sigma}^+ - \frac{1}{2}\{\hat{\sigma}^+\hat{\sigma}^-,\hat{\rho}_{\rm S}\}\right]$, where $\hat{\rho}_{\rm S}$ is the system density operator, and
\begin{equation}
    \gamma(\omega_0) = \frac{2}{\epsilon_0} \mathbf{d} \cdot \text{Im}\{\mathbf{G}_{\perp}(\mathbf{r}_0,\mathbf{r}_0,\omega_0) \} \cdot \mathbf{d},
\end{equation}
and here the Green's function is transverse with respect to both spatial arguments and sides of the dyad.

We can simplify this expression for a situation where only one resonant mode is dominant at the dipole location and frequency by using a QNM expansion of the Green's 
function. The QNMs are solutions to the Helmholtz equation with open boundary conditions (i.e., the Silver-M{\"u}ller radiation condition):
\begin{equation}
    {\bm \nabla} \times {\bm \nabla} \times \tilde{\mathbf{f}}_{\mu}(\mathbf{r}) - \left(\frac{\tilde{\omega}_{\mu}}{c}\right)^2 \epsilon(\mathbf{r},\tilde{\omega}_{\mu})\tilde{\mathbf{f}}_{\mu}(\mathbf{r})=0,
\end{equation}
where $\tilde{\omega}_\mu = \omega_{\mu} - i\kappa_{\mu}/2$ is a complex QNM frequency, with $\kappa_{\mu}$ corresponding to the cavity photon decay rate. The transverse part of the Green's function can be expanded in terms of the QNMs as
$
    \mathbf{G}_{\perp}(\mathbf{r}_0,\mathbf{r}_0,\omega) = \sum_{\mu} A_{\mu}(\omega)\tilde{\mathbf{f}}_{\mu}(\mathbf{r}_0)\tilde{\mathbf{f}}_{\mu}(\mathbf{r}_0),
$
provided that $\mathbf{r}_0$ is sufficiently close to the resonator, and we  use the expansion coefficient, $A_{\mu}(\omega) = \omega/[2(\tilde{\omega}_\mu - \omega)]$ \cite{Kristensen2020Sep}.

Using a single QNM expansion with $\mu = c$, we obtain
\begin{equation}\label{eq:gamma_0}
\gamma(\omega_0) = \frac{4|\tilde{g}_{\rm d}|^2}{\kappa_c}\frac{\omega_0}{\omega_c}\frac{\kappa_c^2/4}{\kappa_c^2/4 + (\omega_0-\omega_c)^2}\chi_c(\phi_0,\omega_0),
\end{equation}
where we have defined
$    \tilde{g}_{\rm d} = \sqrt{\frac{\omega_c}{2\epsilon_0}}\mathbf{d}\cdot\tilde{\mathbf{f}}_c(\mathbf{r}_0)$,
and
\begin{equation}
    \chi_c(\phi_0,\omega_0) = \cos{\left(2\phi_0\right)} - 2Q_c\sin{\left(2\phi_0\right)}\left[\omega_0/\omega_c-1\right]
\end{equation}
is a factor accounting for the phase of the QNM at the position of the dipole, $\phi_0 = \text{arg}\{\tilde{\mathbf{f}}_c(\mathbf{r}_0)\}$\red{, and $Q_c = \omega_c/\kappa_c$}; note $\chi_c =1$ {\it only} when the QNM is real at the dipole location.

For a real-valued QNM at the dipole location, the SE rate takes the form of a Lorentzian function centered around the cavity frequency, modulated by a factor of $\omega_0/\omega_c$. This is in contrast to the phenomenological result often employed, where the decay rate is simply Lorentzian.
However, even for a small non-zero QNM phase at the dipole location, the deviations from a pure Lorentzian can be more complex; for example, expanding to leading order in $\delta=(\omega_0 - \omega_c)/\omega_c$ (excluding the Lorentzian factor), the rate becomes 
\begin{equation}
    \frac{\gamma(\omega_0)}{\gamma(\omega_c)} = \frac{\kappa_c^2/4}{\kappa_c^2/4 + (\omega_0-\omega_c)^2}\left[1+\delta(1-2Q_c\tan{(2\phi_0)}\right],
\end{equation}
and so the QNM phase is only negligible when $4Q_c|\phi_0| \ll 1$, which is a very strict condition.
This derivation can also be carried out in the dipole gauge~\cite{SI}.

\begin{figure}[hbt!]
 \includegraphics[width=0.48\textwidth]{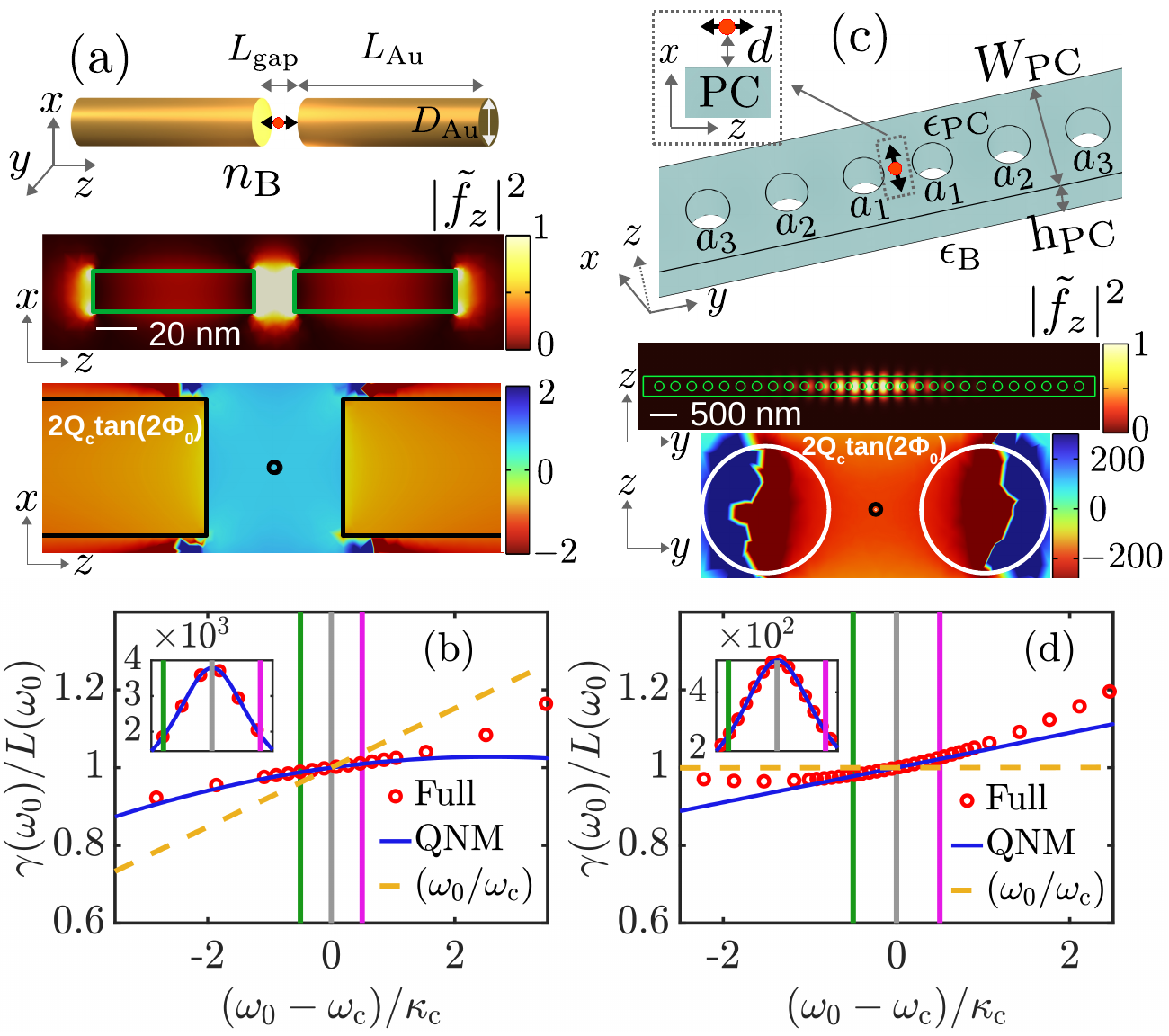}
    \caption{(a) Schematic diagram, QNM profile $|\tilde{f}_{z}|^2$ (dominant $z$ component, arbitrary units) and distribution of $2Q_c\tan{(2\phi_0)}$ of a gold-like dimer. An example dipole location is shown at gap center, where $2Q_c\tan{(2\phi_0)}\approx0.72$ (black circle). 
    (b) The normalized decay rate $\gamma(\omega_0)/L(\omega_0)$ at gap center
    using a single QNM approximation (solid blue curve) and full dipole results (red circles). The dashed orange line shows $\omega_0/\omega_c$. Three vertical lines show the positions of $[-0.5, 0, 0.5]$.
    The inset shows the enhanced decay rate $\gamma(\omega_0)/\gamma_0(\omega_c)$ (Purcell factors), where the single QNM results match perfectly with full dipole results. (c,d) Similar to (a,b), but with a PC beam cavity. 
    The QNM distribution is at the PC beam center.
    Distribution of $2Q_c\tan{(2\phi_0)}$ is at a surface $5~$nm above the PC beam. At the dipole location (black circle), $2Q_c\tan{(2\phi_0)} \approx-190$.
     }
     \label{fig:dimer_PC}
 \end{figure}

In Fig.~\ref{fig:dimer_PC}(b,d), we plot the decay rate $\gamma(\omega_0)$ for two example physical modes corresponding to a metallic dimer (a) and photonic crystal (PC) (c) cavity, normalized by the Lorentzian factor $L(\omega_0)=\gamma(\omega_c)\kappa_c^2/4\left[\kappa_c^2/4 + (\omega_0-\omega_c)^2\right]^{-1}$ in order to investigate the accuracy of the QNM model of the spectral density, compared with a full finite-element 
Green's function simulation with no modal approximations. These calculations are for full 3-D geometries and also include material dispersion and absorption. Near the QNM resonance, the trend agrees well with Eq.~\eqref{eq:gamma_0} evaluated with a single QNM. Further away, deviations arise which can be associated to multi-mode effects (which are easily treated by adding more modes to our formalism; see SM~\cite{SI}).
In contrast, neglecting the QNM phase [which leads to the linear $\sim \omega_0$ scaling modulating $L(\omega_0)$\red{, shown as a dashed yellow line in Fig.~\ref{fig:dimer_PC}(b,d)}] does not accurately reflect the trend.

\emph{Quantized Lossy Cavity Mode Theory.}
We now model our system as a single quantized cavity mode (coupled to a broadband photon reservoir) weakly coupled to a TLS, which is described by the Jaynes-Cummings (JC) model \red{(in the SM~\cite{SI}, we show that the rotating-wave approximation made in the JC model does not affect our results)}. It is sufficient to consider the system as initially in the excited state of the TLS, with zero cavity photons present. 
Under these assumptions, and working in an appropriate rotating frame, this system is described by the Hamiltonian $\hat{H}^{\rm g}(t) = \hat{H}^{\rm g}_{\rm S} + \hat{H}_{\rm int}(t)$, where
\begin{equation}
    \hat{H}_{\rm S}^{\rm g} = \Delta \hat{\sigma}^+ \hat{\sigma}^- + g_{\rm g}\left(\hat{a}\hat{\sigma}^+ + \hat{a}^{\dagger} \hat{\sigma}^-\right),
\end{equation}
and 
\begin{equation}\label{eq:sys-res}
\hat{H}_{\rm int}(t) = \int \! d\omega \Lambda(\omega)\left[\hat{a}^{\dagger} \hat{b}_{\omega}e^{-i(\omega-\omega_c) t} + \hat{a}\hat{b}^{\dagger}_{\omega}e^{i(\omega-\omega_c) t}\right].
\end{equation}
Here, $\hat{a}$, $\hat{a}^{\dagger}$ are cavity mode operators, $\Delta = \omega_0 - \omega_c$, and $g_{\rm g}$ is a gauge-dependent coupling rate. The operators $\hat{b}_{\omega}$ and $\hat{b}_{\omega}^{\dagger}$ are continuum reservoir operators that satisfy $[\hat{b}_{\omega},\hat{b}_{\omega'}^{\dagger}]=\delta(\omega-\omega')$, and we 
let 
$\Lambda(\omega_c)= \sqrt{\kappa_c/(2\pi)}$ to recover an empty cavity photon decay rate $\kappa_c$.

Considering now the gauge-dependence of the model, if using the dipole gauge ($\text{g}=\text{d}$), then $g_{\rm g} \rightarrow |\tilde{g}_{\rm d}|$, and 
in contrast to the continuum model, the phase of $\tilde{\mathbf{f}}_c(\mathbf{r}_0)$ plays no role in the quantized cavity theory presented here, and thus we choose $g_{\rm d} \equiv |\tilde{g}_{\rm d}|$ to be real in this model without loss of generality (i.e., by absorbing the phase of the QNM at the dipole location into the TLS operators by a unitary rotation). 
In general, this can be derived from a \emph{discrete mode projection} from the underlying continuum model described previously, and can also be directly related to quantized QNMs~\cite{Gustin2023Jan,franke_fluctuation-dissipation_2020}.  In the Coulomb gauge, $g_{\rm c} = \frac{\omega_0}{\omega_c}g_{\rm d}$. Since only $g_{\rm d}$ is independent of $\omega_0$, to freely vary the TLS frequency $\omega_0$ independently of the cavity coupling parameters, we express all results in terms of $g_{\rm d}$.

We now diagonalize the system Hamiltonian. From the initial condition, the system is closed under the eigenstates $\ket{G} = \ket{0,g}$ (see SM~\cite{SI} for full model), and 
$
    \ket{\pm} = \frac{1}{\sqrt{2}}\left[\pm \sqrt{1 \pm \frac{\Delta}{\eta}}\ket{0,e} + \sqrt{1 \mp \frac{\Delta}{\eta}}\ket{1,g}    \right]
    $
with the corresponding energies,  $
    \omega_{\pm}  = \frac{\Delta}{2} \pm \frac{1}{2}\eta$,
where $\eta = \sqrt{\Delta^2+4g_{\rm g}^2}$. We decompose the Hamiltonian in terms of these eigenstates as
$
    \hat{H}_{\rm S}^{\rm g} = \sum_{\alpha=\{+,-\}}\omega_{\alpha} \ket{\alpha}\bra{\alpha},
    $
and
\begin{equation}
    \hat{H}_{\rm int}(t) = \sum_{\alpha =\{+,-\}} \! \!c_{\alpha}\ket{G}\bra{\alpha} \int d\omega \Lambda(\omega) \hat{b}_{\omega}^{\dagger} e^{i(\omega-\omega_c)t} + \text{H.c.},
\end{equation}
where $c_{\alpha} = \sqrt{\left(1-\alpha \Delta/\eta\right)/2}$.

Following a 2nd-order Born-Markov approximation, the master equation for the system is
    $\dot{\hat{\rho}}_{\rm S} = -i[\hat{H}^{\rm g}_{\rm S},\hat{\rho}_{\rm S}] + 
    \int_0^{\infty} \! \! d\tau\text{Tr}_{\rm R}\left([e^{-i \hat{H}_{\rm S}^{\rm g}\tau}\hat{H}_{\rm int}(t\!-\!\tau) e^{i\hat{H}_{\rm S}^{\rm g}\tau}\hat{\rho}_{\rm S} \hat{\rho}_{\rm R},\hat{H}_{\rm int}(t)]\right] + \text{H.c.}$,
which evaluates to
\begin{align}\label{eq:me_full}
    \dot{\hat{\rho}}_{\rm S} = -i[\hat{H}^{\rm g}_{\rm S},\hat{\rho}_{\rm S}]+&  \pi\sum_{\alpha}c_{\alpha}\Lambda^2(\omega_c+\omega_{\alpha})\Big( \nonumber \\ 
    & \Big[\ket{G}\bra{\alpha} \hat{\rho}_{\rm S}, \hat{a}^{\dagger}\Big] + \text{H.c.} \Big).
\end{align}

To determine the TLS decay rate $\gamma$, we can solve for the eigenvalues of Eq.~\eqref{eq:me_full}, viewed as a matrix equation for the elements of $\hat{\rho}_{\rm S}$. 
For the weak-coupling decay rate, we expect $\gamma \sim g_{\rm d}^2$, from the usual Purcell effect. Thus, we 
consider the eigenvalue $\gamma_{\rm g} \sim g_{\rm g}^2$, and use perturbation theory (see~\cite{SI} Sec. III)
to find, to second-order in $g_{\rm g}$:
\begin{equation}
    \gamma_{\rm g}(\omega_0) = g_{\rm g}^2\frac{2\pi \Lambda^2(\omega_0)}{\kappa_c^2/4 + \Delta^2}. 
\end{equation}
Now we employ the ansatz
\begin{equation}\label{eq:spectralfn}
    \Lambda(\omega) = \sqrt{\frac{\kappa_c}{2\pi}}\left(\frac{\omega}{\omega_c}\right)^n,
\end{equation}
\begin{equation}\label{eq:gamma_gauge}
   \rightarrow  \gamma_{\rm g} = \left(\frac{4g_{\rm g}^2}{\kappa_c}\right)\left(\frac{\omega_0}{\omega_c}\right)^{2n}\left[\frac{\kappa_c^2/4}{\kappa_c^2/4 + \Delta^2}\right].
\end{equation}

If
working in the dipole gauge, then we must choose $n = \frac{1}{2}$ to satisfy Eq.~\eqref{eq:gamma_0}, for the case of a \emph{real} valued QNM. In contrast, in the Coulomb gauge, $g_{\rm c} = \frac{\omega_0}{\omega_c}g_{\rm d}$, and so we must choose $n=-1/2$.  %
\red{If we choose the Coulomb gauge,} we see that 
\begin{equation}
    \Lambda_{\rm C}(\omega) = \sqrt{\left(\frac{\kappa_c}{2\pi}\right)\left(\frac{\omega_c}{\omega}\right)} 
\end{equation}
is the system-bath coupling function which, in cavity-QED, produces the correct results for an isolated mode with a strictly real QNM at the dipole location; 
surprisingly, this corresponds to a \emph{sub-Ohmic} reservoir (in contrast, say, to the Ohmic result encountered in circuit-QED~\cite{Blais2021May}, \red{ sometimes} also assumed 
in optics). \red{In contrast, if we choose the dipole gauge, then the system-bath coupling function consistent with Eq.~\eqref{eq:sys-res} is
\begin{equation}
    \Lambda_{\rm d}(\omega) = \sqrt{\left(\frac{\kappa_c}{2\pi}\right)\left(\frac{\omega}{\omega_c}\right)}. 
\end{equation}
We stress that the different choice of spectral density specifically corresponds to which gauge has a coupling Hamiltonian of the form in Eq.~\eqref{eq:sys-res}, where the reservoir couples to bosonic cavity operators. In the SM~\cite{SI}, we show that if the operators in the interaction Hamiltonian are also transformed with the gauge transformation, along with the system Hamiltonian, then the decay rate $\gamma_{\rm g}$ is the same in either gauge, for a fixed spectral density. 
} \red{We note that these spectral densities are \emph{local forms}, valid in a spectral range where other modes of the cavity are not appreciable.}

The analytic result derived in Eq.~\eqref{eq:gamma_gauge} for the Coulomb gauge [such that it is equivalent to Eq.~\eqref{eq:gamma_0}] agrees excellently with a full numerical calculation from the master equation in Eq.~\eqref{eq:me_full}, which we show in the
Supplementary Material (SM)~\cite{SI}.
However,
 the case where the QNM is not real at the TLS location is much more realistic for practical mode geometries. In Fig.~\ref{fig:dimer_PC} (a,c), we show the QNM phase $\phi(\mathbf{r})$ as a function of position for the two considered example physical modes.
The product $2Q_c\tan{(2\phi_0)}$ remains substantial at all locations near the modal electric field maximum (even where $|\phi_0| \ll 1$).

 \begin{figure}[th]
 \includegraphics[width=0.46\textwidth]{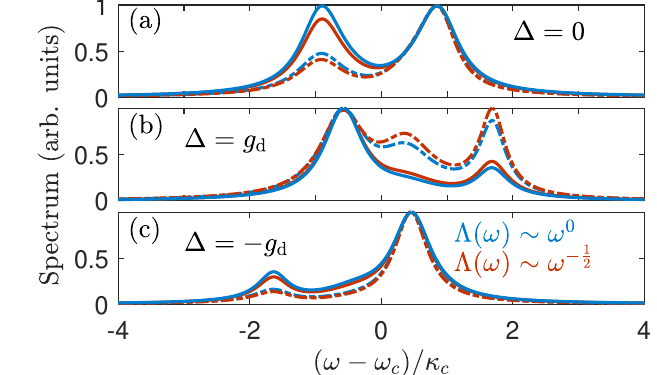}
    \caption{
    Normalized cavity 
    emission spectrum under incoherent excitation calculated from the master equation in Eq.~\eqref{eq:me_full} in the strong-coupling regime \red{using the Coulomb gauge}. Here we consider a cavity with $Q_c=20$, $g_{\rm d}/\omega_c = 0.05$, and $\phi_0=0.03$. The solid lines correspond to a spectral density with $\Lambda(\omega) = \sqrt{\frac{\kappa}{2\pi}}\left(\omega/\omega_c\right)^n$, where $n=0$ (blue) or $n = -\frac{1}{2}$ (red), and the dash-dotted lines correspond to $\Lambda(\omega) = \sqrt{\frac{\kappa}{2\pi}}\left(\omega/\omega_c\right)^n\sqrt{\chi_c(\phi_0,\omega)}$.
    }
     \label{fig:sc}
 \end{figure}

It is difficult if not impossible to obtain the general phase-dependent result in Eq.~\eqref{eq:gamma_0} using simple system-reservoir constructions of a single lossy cavity with a general spectral bath coupling $\Lambda(\omega)$ which is independent of the contents of the cavity. This points to important and \emph{potentially fundamental} limitations to the ability of these simple universal models of the spectral density to predict accurately observables in the USC regime. We note that other approaches, such as those utilizing rigorous quantized QNM expansions may bypass this limitation. Additionally, one can make the heuristic replacement 
\begin{equation}\label{eq:heuristic}
\Lambda(\omega) \rightarrow \red{\Lambda_{\rm C/d}(\omega)}\sqrt{\chi_c(\phi_0,\omega)},
\end{equation}
 which recovers the correct result, but is clearly not a universal form of the spectral density as it implies the cavity-bath interaction should be dependent on the position of a dipole placed inside the cavity. This function is shown in Fig.~\ref{fig:schematic}(c) for various values of $\phi_0$.

Although a model-specific and ad-hoc fix, this replacement to the system-bath coupling function \emph{does} allow for quantitatively accurate spectral modelling in the \emph{strong-coupling} regime (and with additional assumptions, the USC regime), which has previously been inaccessible using few-mode master equation methods. In Fig.~\ref{fig:sc}, we show simulations highlighting the significance of this replacement,  considering intrinsically quantum effects by pumping the system beyond the weak excitation regime (details in the SM~\cite{SI}). Notably, even outside of the USC regime (where such considerations become even more important), the importance of both the QNM phase-dependent correction factor and the correct coupling function prefactor \red{$\sim \omega^{\pm \frac{1}{2}}$ }can be highly substantial, which has significant implications for current commonly used models, which typically assume a flat or Ohmic reservoir for the cavity. It should be noted that for sufficiently broadband dynamics (e.g., large enough detuning $\Delta$ or cavity-dipole coupling $g_{\rm g}$), $\chi_c$ can become negative (and thus the spectral density becomes unphysical) for certain transitions, which points to potential \emph{fundamental limits} of single-mode models in such regimes. Indeed Fig.~\ref{fig:dimer_PC}(b,d) already shows deviations arising from additional mode contributions.

\emph{Conclusions.}
By comparison with a rigorous gauge-invariant macroscopic QED formalism in arbitrary media, we have shown the correct form of the spectral density for a lossy quantized cavity mode interacting with a TLS. The spectral density takes a different form than those typically assumed in previous works, is not universal (it depends on the position of the TLS), and can significantly impact predictions even in the weak and strong coupling regimes. This form is also gauge-dependent. \red{An interesting extension of this work would be to consider the case of multiple dipoles placed at spatial locations with substantially differing QNM phases.}

In a future work, we will use an \emph{ab initio} construction of the system-reservoir Hamiltonian by means of a quantized QNM approach to study dissipative cavity QED in the USC regime, beyond the phenomenological approaches which are known to fail to predict things like emission spectra accurately~\cite{Salmon2022Mar}, and further investigate the fundamental limits to single-mode models.

This work was supported by the Natural Sciences and Engineering Research Council of Canada (NSERC),
the Canadian Foundation for Innovation (CFI), Queen's University, Canada, NSF awards PHY-2011363 and CCF-1918549, and CMC Microsystems for the provision of COMSOL Multiphysics.
We thank Sebastian Franke and Hideo Mabuchi
for useful discussions and comments.

\bibliography{bib}

\begin{thebibliography}{48}%
\makeatletter
\providecommand \@ifxundefined [1]{%
 \@ifx{#1\undefined}
}%
\providecommand \@ifnum [1]{%
 \ifnum #1\expandafter \@firstoftwo
 \else \expandafter \@secondoftwo
 \fi
}%
\providecommand \@ifx [1]{%
 \ifx #1\expandafter \@firstoftwo
 \else \expandafter \@secondoftwo
 \fi
}%
\providecommand \natexlab [1]{#1}%
\providecommand \enquote  [1]{``#1''}%
\providecommand \bibnamefont  [1]{#1}%
\providecommand \bibfnamefont [1]{#1}%
\providecommand \citenamefont [1]{#1}%
\providecommand \href@noop [0]{\@secondoftwo}%
\providecommand \href [0]{\begingroup \@sanitize@url \@href}%
\providecommand \@href[1]{\@@startlink{#1}\@@href}%
\providecommand \@@href[1]{\endgroup#1\@@endlink}%
\providecommand \@sanitize@url [0]{\catcode `\\12\catcode `\$12\catcode `\&12\catcode `\#12\catcode `\^12\catcode `\_12\catcode `\%12\relax}%
\providecommand \@@startlink[1]{}%
\providecommand \@@endlink[0]{}%
\providecommand \url  [0]{\begingroup\@sanitize@url \@url }%
\providecommand \@url [1]{\endgroup\@href {#1}{\urlprefix }}%
\providecommand \urlprefix  [0]{URL }%
\providecommand \Eprint [0]{\href }%
\providecommand \doibase [0]{https://doi.org/}%
\providecommand \selectlanguage [0]{\@gobble}%
\providecommand \bibinfo  [0]{\@secondoftwo}%
\providecommand \bibfield  [0]{\@secondoftwo}%
\providecommand \translation [1]{[#1]}%
\providecommand \BibitemOpen [0]{}%
\providecommand \bibitemStop [0]{}%
\providecommand \bibitemNoStop [0]{.\EOS\space}%
\providecommand \EOS [0]{\spacefactor3000\relax}%
\providecommand \BibitemShut  [1]{\csname bibitem#1\endcsname}%
\let\auto@bib@innerbib\@empty
\bibitem [{\citenamefont {Giustina}\ \emph {et~al.}(2015)\citenamefont {Giustina}, \citenamefont {Versteegh}, \citenamefont {Wengerowsky}, \citenamefont {Handsteiner}, \citenamefont {Hochrainer}, \citenamefont {Phelan}, \citenamefont {Steinlechner}, \citenamefont {Kofler}, \citenamefont {Larsson}, \citenamefont {Abell{\ifmmode\acute{a}\else\'{a}\fi}n}, \citenamefont {Amaya}, \citenamefont {Pruneri}, \citenamefont {Mitchell}, \citenamefont {Beyer}, \citenamefont {Gerrits}, \citenamefont {Lita}, \citenamefont {Shalm}, \citenamefont {Nam}, \citenamefont {Scheidl}, \citenamefont {Ursin}, \citenamefont {Wittmann},\ and\ \citenamefont {Zeilinger}}]{Giustina2015Dec}%
  \BibitemOpen
  \bibfield  {author} {\bibinfo {author} {\bibfnamefont {M.}~\bibnamefont {Giustina}}, \bibinfo {author} {\bibfnamefont {M.~A.~M.}\ \bibnamefont {Versteegh}}, \bibinfo {author} {\bibfnamefont {S.}~\bibnamefont {Wengerowsky}}, \bibinfo {author} {\bibfnamefont {J.}~\bibnamefont {Handsteiner}}, \bibinfo {author} {\bibfnamefont {A.}~\bibnamefont {Hochrainer}}, \bibinfo {author} {\bibfnamefont {K.}~\bibnamefont {Phelan}}, \bibinfo {author} {\bibfnamefont {F.}~\bibnamefont {Steinlechner}}, \bibinfo {author} {\bibfnamefont {J.}~\bibnamefont {Kofler}}, \bibinfo {author} {\bibfnamefont {J.-{\AA}.}\ \bibnamefont {Larsson}}, \bibinfo {author} {\bibfnamefont {C.}~\bibnamefont {Abell{\ifmmode\acute{a}\else\'{a}\fi}n}}, \bibinfo {author} {\bibfnamefont {W.}~\bibnamefont {Amaya}}, \bibinfo {author} {\bibfnamefont {V.}~\bibnamefont {Pruneri}}, \bibinfo {author} {\bibfnamefont {M.~W.}\ \bibnamefont {Mitchell}}, \bibinfo {author} {\bibfnamefont {J.}~\bibnamefont {Beyer}}, \bibinfo {author} {\bibfnamefont {T.}~\bibnamefont
  {Gerrits}}, \bibinfo {author} {\bibfnamefont {A.~E.}\ \bibnamefont {Lita}}, \bibinfo {author} {\bibfnamefont {L.~K.}\ \bibnamefont {Shalm}}, \bibinfo {author} {\bibfnamefont {S.~W.}\ \bibnamefont {Nam}}, \bibinfo {author} {\bibfnamefont {T.}~\bibnamefont {Scheidl}}, \bibinfo {author} {\bibfnamefont {R.}~\bibnamefont {Ursin}}, \bibinfo {author} {\bibfnamefont {B.}~\bibnamefont {Wittmann}},\ and\ \bibinfo {author} {\bibfnamefont {A.}~\bibnamefont {Zeilinger}},\ }\bibfield  {title} {\bibinfo {title} {{Significant-Loophole-Free Test of Bell's Theorem with Entangled Photons}},\ }\href {https://doi.org/10.1103/PhysRevLett.115.250401} {\bibfield  {journal} {\bibinfo  {journal} {Phys. Rev. Lett.}\ }\textbf {\bibinfo {volume} {115}},\ \bibinfo {pages} {250401} (\bibinfo {year} {2015})}\BibitemShut {NoStop}%
\bibitem [{\citenamefont {Shalm}\ \emph {et~al.}(2015)\citenamefont {Shalm}, \citenamefont {Meyer-Scott}, \citenamefont {Christensen}, \citenamefont {Bierhorst}, \citenamefont {Wayne}, \citenamefont {Stevens}, \citenamefont {Gerrits}, \citenamefont {Glancy}, \citenamefont {Hamel}, \citenamefont {Allman}, \citenamefont {Coakley}, \citenamefont {Dyer}, \citenamefont {Hodge}, \citenamefont {Lita}, \citenamefont {Verma}, \citenamefont {Lambrocco}, \citenamefont {Tortorici}, \citenamefont {Migdall}, \citenamefont {Zhang}, \citenamefont {Kumor}, \citenamefont {Farr}, \citenamefont {Marsili}, \citenamefont {Shaw}, \citenamefont {Stern}, \citenamefont {Abell{\ifmmode\acute{a}\else\'{a}\fi}n}, \citenamefont {Amaya}, \citenamefont {Pruneri}, \citenamefont {Jennewein}, \citenamefont {Mitchell}, \citenamefont {Kwiat}, \citenamefont {Bienfang}, \citenamefont {Mirin}, \citenamefont {Knill},\ and\ \citenamefont {Nam}}]{Shalm2015Dec}%
  \BibitemOpen
  \bibfield  {author} {\bibinfo {author} {\bibfnamefont {L.~K.}\ \bibnamefont {Shalm}}, \bibinfo {author} {\bibfnamefont {E.}~\bibnamefont {Meyer-Scott}}, \bibinfo {author} {\bibfnamefont {B.~G.}\ \bibnamefont {Christensen}}, \bibinfo {author} {\bibfnamefont {P.}~\bibnamefont {Bierhorst}}, \bibinfo {author} {\bibfnamefont {M.~A.}\ \bibnamefont {Wayne}}, \bibinfo {author} {\bibfnamefont {M.~J.}\ \bibnamefont {Stevens}}, \bibinfo {author} {\bibfnamefont {T.}~\bibnamefont {Gerrits}}, \bibinfo {author} {\bibfnamefont {S.}~\bibnamefont {Glancy}}, \bibinfo {author} {\bibfnamefont {D.~R.}\ \bibnamefont {Hamel}}, \bibinfo {author} {\bibfnamefont {M.~S.}\ \bibnamefont {Allman}}, \bibinfo {author} {\bibfnamefont {K.~J.}\ \bibnamefont {Coakley}}, \bibinfo {author} {\bibfnamefont {S.~D.}\ \bibnamefont {Dyer}}, \bibinfo {author} {\bibfnamefont {C.}~\bibnamefont {Hodge}}, \bibinfo {author} {\bibfnamefont {A.~E.}\ \bibnamefont {Lita}}, \bibinfo {author} {\bibfnamefont {V.~B.}\ \bibnamefont {Verma}}, \bibinfo {author}
  {\bibfnamefont {C.}~\bibnamefont {Lambrocco}}, \bibinfo {author} {\bibfnamefont {E.}~\bibnamefont {Tortorici}}, \bibinfo {author} {\bibfnamefont {A.~L.}\ \bibnamefont {Migdall}}, \bibinfo {author} {\bibfnamefont {Y.}~\bibnamefont {Zhang}}, \bibinfo {author} {\bibfnamefont {D.~R.}\ \bibnamefont {Kumor}}, \bibinfo {author} {\bibfnamefont {W.~H.}\ \bibnamefont {Farr}}, \bibinfo {author} {\bibfnamefont {F.}~\bibnamefont {Marsili}}, \bibinfo {author} {\bibfnamefont {M.~D.}\ \bibnamefont {Shaw}}, \bibinfo {author} {\bibfnamefont {J.~A.}\ \bibnamefont {Stern}}, \bibinfo {author} {\bibfnamefont {C.}~\bibnamefont {Abell{\ifmmode\acute{a}\else\'{a}\fi}n}}, \bibinfo {author} {\bibfnamefont {W.}~\bibnamefont {Amaya}}, \bibinfo {author} {\bibfnamefont {V.}~\bibnamefont {Pruneri}}, \bibinfo {author} {\bibfnamefont {T.}~\bibnamefont {Jennewein}}, \bibinfo {author} {\bibfnamefont {M.~W.}\ \bibnamefont {Mitchell}}, \bibinfo {author} {\bibfnamefont {P.~G.}\ \bibnamefont {Kwiat}}, \bibinfo {author} {\bibfnamefont {J.~C.}\
  \bibnamefont {Bienfang}}, \bibinfo {author} {\bibfnamefont {R.~P.}\ \bibnamefont {Mirin}}, \bibinfo {author} {\bibfnamefont {E.}~\bibnamefont {Knill}},\ and\ \bibinfo {author} {\bibfnamefont {S.~W.}\ \bibnamefont {Nam}},\ }\bibfield  {title} {\bibinfo {title} {{Strong Loophole-Free Test of Local Realism}},\ }\href {https://doi.org/10.1103/PhysRevLett.115.250402} {\bibfield  {journal} {\bibinfo  {journal} {Phys. Rev. Lett.}\ }\textbf {\bibinfo {volume} {115}},\ \bibinfo {pages} {250402} (\bibinfo {year} {2015})}\BibitemShut {NoStop}%
\bibitem [{\citenamefont {Mabuchi}\ and\ \citenamefont {Doherty}(2002)}]{Mabuchi2002Nov}%
  \BibitemOpen
  \bibfield  {author} {\bibinfo {author} {\bibfnamefont {H.}~\bibnamefont {Mabuchi}}\ and\ \bibinfo {author} {\bibfnamefont {A.~C.}\ \bibnamefont {Doherty}},\ }\bibfield  {title} {\bibinfo {title} {{Cavity Quantum Electrodynamics: Coherence in Context}},\ }\href {https://doi.org/10.1126/science.1078446} {\bibfield  {journal} {\bibinfo  {journal} {Science}\ }\textbf {\bibinfo {volume} {298}},\ \bibinfo {pages} {1372} (\bibinfo {year} {2002})}\BibitemShut {NoStop}%
\bibitem [{\citenamefont {Weedbrook}\ \emph {et~al.}(2012)\citenamefont {Weedbrook}, \citenamefont {Pirandola}, \citenamefont {Garc{\ifmmode\acute{\imath}\else\'{\i}\fi}a-Patr{\ifmmode\acute{o}\else\'{o}\fi}n}, \citenamefont {Cerf}, \citenamefont {Ralph}, \citenamefont {Shapiro},\ and\ \citenamefont {Lloyd}}]{Weedbrook2012May}%
  \BibitemOpen
  \bibfield  {author} {\bibinfo {author} {\bibfnamefont {C.}~\bibnamefont {Weedbrook}}, \bibinfo {author} {\bibfnamefont {S.}~\bibnamefont {Pirandola}}, \bibinfo {author} {\bibfnamefont {R.}~\bibnamefont {Garc{\ifmmode\acute{\imath}\else\'{\i}\fi}a-Patr{\ifmmode\acute{o}\else\'{o}\fi}n}}, \bibinfo {author} {\bibfnamefont {N.~J.}\ \bibnamefont {Cerf}}, \bibinfo {author} {\bibfnamefont {T.~C.}\ \bibnamefont {Ralph}}, \bibinfo {author} {\bibfnamefont {J.~H.}\ \bibnamefont {Shapiro}},\ and\ \bibinfo {author} {\bibfnamefont {S.}~\bibnamefont {Lloyd}},\ }\bibfield  {title} {\bibinfo {title} {{Gaussian quantum information}},\ }\href {https://doi.org/10.1103/RevModPhys.84.621} {\bibfield  {journal} {\bibinfo  {journal} {Rev. Mod. Phys.}\ }\textbf {\bibinfo {volume} {84}},\ \bibinfo {pages} {621} (\bibinfo {year} {2012})}\BibitemShut {NoStop}%
\bibitem [{\citenamefont {Browne}\ \emph {et~al.}(2017)\citenamefont {Browne}, \citenamefont {Bose}, \citenamefont {Mintert},\ and\ \citenamefont {Kim}}]{Browne2017}%
  \BibitemOpen
  \bibfield  {author} {\bibinfo {author} {\bibfnamefont {D.}~\bibnamefont {Browne}}, \bibinfo {author} {\bibfnamefont {S.}~\bibnamefont {Bose}}, \bibinfo {author} {\bibfnamefont {F.}~\bibnamefont {Mintert}},\ and\ \bibinfo {author} {\bibfnamefont {M.}~\bibnamefont {Kim}},\ }\bibfield  {title} {\bibinfo {title} {From quantum optics to quantum technologies},\ }\href {https://doi.org/10.1016/j.pquantelec.2017.06.002} {\bibfield  {journal} {\bibinfo  {journal} {Progress in Quantum Electronics}\ }\textbf {\bibinfo {volume} {54}},\ \bibinfo {pages} {2–18} (\bibinfo {year} {2017})}\BibitemShut {NoStop}%
\bibitem [{\citenamefont {Lambropoulos}\ \emph {et~al.}(2000)\citenamefont {Lambropoulos}, \citenamefont {Nikolopoulos}, \citenamefont {Nielsen},\ and\ \citenamefont {Bay}}]{Lambropoulos2000}%
  \BibitemOpen
  \bibfield  {author} {\bibinfo {author} {\bibfnamefont {P.}~\bibnamefont {Lambropoulos}}, \bibinfo {author} {\bibfnamefont {G.~M.}\ \bibnamefont {Nikolopoulos}}, \bibinfo {author} {\bibfnamefont {T.~R.}\ \bibnamefont {Nielsen}},\ and\ \bibinfo {author} {\bibfnamefont {S.}~\bibnamefont {Bay}},\ }\bibfield  {title} {\bibinfo {title} {Fundamental quantum optics in structured reservoirs},\ }\href {https://doi.org/10.1088/0034-4885/63/4/201} {\bibfield  {journal} {\bibinfo  {journal} {Reports on Progress in Physics}\ }\textbf {\bibinfo {volume} {63}},\ \bibinfo {pages} {455–503} (\bibinfo {year} {2000})}\BibitemShut {NoStop}%
\bibitem [{\citenamefont {Reiserer}\ and\ \citenamefont {Rempe}(2015)}]{RevModPhys.87.1379}%
  \BibitemOpen
  \bibfield  {author} {\bibinfo {author} {\bibfnamefont {A.}~\bibnamefont {Reiserer}}\ and\ \bibinfo {author} {\bibfnamefont {G.}~\bibnamefont {Rempe}},\ }\bibfield  {title} {\bibinfo {title} {Cavity-based quantum networks with single atoms and optical photons},\ }\href {https://doi.org/10.1103/RevModPhys.87.1379} {\bibfield  {journal} {\bibinfo  {journal} {Rev. Mod. Phys.}\ }\textbf {\bibinfo {volume} {87}},\ \bibinfo {pages} {1379} (\bibinfo {year} {2015})}\BibitemShut {NoStop}%
\bibitem [{\citenamefont {Lodahl}\ \emph {et~al.}(2015)\citenamefont {Lodahl}, \citenamefont {Mahmoodian},\ and\ \citenamefont {Stobbe}}]{RevModPhys.87.347}%
  \BibitemOpen
  \bibfield  {author} {\bibinfo {author} {\bibfnamefont {P.}~\bibnamefont {Lodahl}}, \bibinfo {author} {\bibfnamefont {S.}~\bibnamefont {Mahmoodian}},\ and\ \bibinfo {author} {\bibfnamefont {S.}~\bibnamefont {Stobbe}},\ }\bibfield  {title} {\bibinfo {title} {Interfacing single photons and single quantum dots with photonic nanostructures},\ }\href {https://doi.org/10.1103/RevModPhys.87.347} {\bibfield  {journal} {\bibinfo  {journal} {Rev. Mod. Phys.}\ }\textbf {\bibinfo {volume} {87}},\ \bibinfo {pages} {347} (\bibinfo {year} {2015})}\BibitemShut {NoStop}%
\bibitem [{\citenamefont {Haroche}\ \emph {et~al.}(2020)\citenamefont {Haroche}, \citenamefont {Brune},\ and\ \citenamefont {Raimond}}]{Haroche2020}%
  \BibitemOpen
  \bibfield  {author} {\bibinfo {author} {\bibfnamefont {S.}~\bibnamefont {Haroche}}, \bibinfo {author} {\bibfnamefont {M.}~\bibnamefont {Brune}},\ and\ \bibinfo {author} {\bibfnamefont {J.~M.}\ \bibnamefont {Raimond}},\ }\bibfield  {title} {\bibinfo {title} {From cavity to circuit quantum electrodynamics},\ }\href {https://doi.org/10.1038/s41567-020-0812-1} {\bibfield  {journal} {\bibinfo  {journal} {Nature Physics}\ }\textbf {\bibinfo {volume} {16}},\ \bibinfo {pages} {243–246} (\bibinfo {year} {2020})}\BibitemShut {NoStop}%
\bibitem [{\citenamefont {Yurke}\ and\ \citenamefont {Denker}(1984)}]{Yurke1984Mar}%
  \BibitemOpen
  \bibfield  {author} {\bibinfo {author} {\bibfnamefont {B.}~\bibnamefont {Yurke}}\ and\ \bibinfo {author} {\bibfnamefont {J.~S.}\ \bibnamefont {Denker}},\ }\bibfield  {title} {\bibinfo {title} {{Quantum network theory}},\ }\href {https://doi.org/10.1103/PhysRevA.29.1419} {\bibfield  {journal} {\bibinfo  {journal} {Phys. Rev. A}\ }\textbf {\bibinfo {volume} {29}},\ \bibinfo {pages} {1419} (\bibinfo {year} {1984})}\BibitemShut {NoStop}%
\bibitem [{\citenamefont {Gardiner}\ and\ \citenamefont {Collett}(1985)}]{Gardiner1985Jun}%
  \BibitemOpen
  \bibfield  {author} {\bibinfo {author} {\bibfnamefont {C.~W.}\ \bibnamefont {Gardiner}}\ and\ \bibinfo {author} {\bibfnamefont {M.~J.}\ \bibnamefont {Collett}},\ }\bibfield  {title} {\bibinfo {title} {{Input and output in damped quantum systems: Quantum stochastic differential equations and the master equation}},\ }\href {https://doi.org/10.1103/PhysRevA.31.3761} {\bibfield  {journal} {\bibinfo  {journal} {Phys. Rev. A}\ }\textbf {\bibinfo {volume} {31}},\ \bibinfo {pages} {3761} (\bibinfo {year} {1985})}\BibitemShut {NoStop}%
\bibitem [{\citenamefont {Frisk~Kockum}\ \emph {et~al.}(2019)\citenamefont {Frisk~Kockum}, \citenamefont {Miranowicz}, \citenamefont {De~Liberato}, \citenamefont {Savasta},\ and\ \citenamefont {Nori}}]{frisk_kockum_ultrastrong_2019}%
  \BibitemOpen
  \bibfield  {author} {\bibinfo {author} {\bibfnamefont {A.}~\bibnamefont {Frisk~Kockum}}, \bibinfo {author} {\bibfnamefont {A.}~\bibnamefont {Miranowicz}}, \bibinfo {author} {\bibfnamefont {S.}~\bibnamefont {De~Liberato}}, \bibinfo {author} {\bibfnamefont {S.}~\bibnamefont {Savasta}},\ and\ \bibinfo {author} {\bibfnamefont {F.}~\bibnamefont {Nori}},\ }\bibfield  {title} {\bibinfo {title} {Ultrastrong coupling between light and matter},\ }\href {https://doi.org/10.1038/s42254-018-0006-2} {\bibfield  {journal} {\bibinfo  {journal} {Nature Reviews Physics}\ }\textbf {\bibinfo {volume} {1}},\ \bibinfo {pages} {19} (\bibinfo {year} {2019})}\BibitemShut {NoStop}%
\bibitem [{\citenamefont {Forn-D{\'i}az}\ \emph {et~al.}(2019)\citenamefont {Forn-D{\'i}az}, \citenamefont {Lamata}, \citenamefont {Rico}, \citenamefont {Kono},\ and\ \citenamefont {Solano}}]{forn-diaz_ultrastrong_2019}%
  \BibitemOpen
  \bibfield  {author} {\bibinfo {author} {\bibfnamefont {P.}~\bibnamefont {Forn-D{\'i}az}}, \bibinfo {author} {\bibfnamefont {L.}~\bibnamefont {Lamata}}, \bibinfo {author} {\bibfnamefont {E.}~\bibnamefont {Rico}}, \bibinfo {author} {\bibfnamefont {J.}~\bibnamefont {Kono}},\ and\ \bibinfo {author} {\bibfnamefont {E.}~\bibnamefont {Solano}},\ }\bibfield  {title} {\bibinfo {title} {Ultrastrong coupling regimes of light-matter interaction},\ }\href {https://doi.org/10.1103/RevModPhys.91.025005} {\bibfield  {journal} {\bibinfo  {journal} {Reviews of Modern Physics}\ }\textbf {\bibinfo {volume} {91}},\ \bibinfo {pages} {025005} (\bibinfo {year} {2019})}\BibitemShut {NoStop}%
\bibitem [{\citenamefont {Salmon}\ \emph {et~al.}(2022)\citenamefont {Salmon}, \citenamefont {Gustin}, \citenamefont {Settineri}, \citenamefont {Di~Stefano}, \citenamefont {Zueco}, \citenamefont {Savasta}, \citenamefont {Nori},\ and\ \citenamefont {Hughes}}]{Salmon2022Mar}%
  \BibitemOpen
  \bibfield  {author} {\bibinfo {author} {\bibfnamefont {W.}~\bibnamefont {Salmon}}, \bibinfo {author} {\bibfnamefont {C.}~\bibnamefont {Gustin}}, \bibinfo {author} {\bibfnamefont {A.}~\bibnamefont {Settineri}}, \bibinfo {author} {\bibfnamefont {O.}~\bibnamefont {Di~Stefano}}, \bibinfo {author} {\bibfnamefont {D.}~\bibnamefont {Zueco}}, \bibinfo {author} {\bibfnamefont {S.}~\bibnamefont {Savasta}}, \bibinfo {author} {\bibfnamefont {F.}~\bibnamefont {Nori}},\ and\ \bibinfo {author} {\bibfnamefont {S.}~\bibnamefont {Hughes}},\ }\bibfield  {title} {\bibinfo {title} {{Gauge-independent emission spectra and quantum correlations in the ultrastrong coupling regime of open system cavity-QED}},\ }\href {https://doi.org/10.1515/nanoph-2021-0718} {\bibfield  {journal} {\bibinfo  {journal} {Nanophotonics}\ }\textbf {\bibinfo {volume} {11}},\ \bibinfo {pages} {1573} (\bibinfo {year} {2022})}\BibitemShut {NoStop}%
\bibitem [{\citenamefont {Hughes}\ \emph {et~al.}(2024)\citenamefont {Hughes}, \citenamefont {Gustin},\ and\ \citenamefont {Nori}}]{Hughes2023Sep}%
  \BibitemOpen
  \bibfield  {author} {\bibinfo {author} {\bibfnamefont {S.}~\bibnamefont {Hughes}}, \bibinfo {author} {\bibfnamefont {C.}~\bibnamefont {Gustin}},\ and\ \bibinfo {author} {\bibfnamefont {F.}~\bibnamefont {Nori}},\ }\bibfield  {title} {\bibinfo {title} {Reconciling quantum and classical spectral theories of ultrastrong coupling: Role of cavity bath coupling and gauge corrections},\ }\href {https://doi.org/10.1364/OPTICAQ.519395} {\bibfield  {journal} {\bibinfo  {journal} {Optica Quantum}\ }\textbf {\bibinfo {volume} {2}},\ \bibinfo {pages} {133} (\bibinfo {year} {2024})}\BibitemShut {NoStop}%
\bibitem [{\citenamefont {Bamba}\ and\ \citenamefont {Ogawa}(2014)}]{Bamba2014Feb}%
  \BibitemOpen
  \bibfield  {author} {\bibinfo {author} {\bibfnamefont {M.}~\bibnamefont {Bamba}}\ and\ \bibinfo {author} {\bibfnamefont {T.}~\bibnamefont {Ogawa}},\ }\bibfield  {title} {\bibinfo {title} {{Recipe for the Hamiltonian of system-environment coupling applicable to the ultrastrong-light-matter-interaction regime}},\ }\href {https://doi.org/10.1103/PhysRevA.89.023817} {\bibfield  {journal} {\bibinfo  {journal} {Phys. Rev. A}\ }\textbf {\bibinfo {volume} {89}},\ \bibinfo {pages} {023817} (\bibinfo {year} {2014})}\BibitemShut {NoStop}%
\bibitem [{\citenamefont {Lentrodt}\ and\ \citenamefont {Evers}(2020)}]{Lentrodt2020Jan}%
  \BibitemOpen
  \bibfield  {author} {\bibinfo {author} {\bibfnamefont {D.}~\bibnamefont {Lentrodt}}\ and\ \bibinfo {author} {\bibfnamefont {J.}~\bibnamefont {Evers}},\ }\bibfield  {title} {\bibinfo {title} {{Ab Initio Few-Mode Theory for Quantum Potential Scattering Problems}},\ }\href {https://doi.org/10.1103/PhysRevX.10.011008} {\bibfield  {journal} {\bibinfo  {journal} {Phys. Rev. X}\ }\textbf {\bibinfo {volume} {10}},\ \bibinfo {pages} {011008} (\bibinfo {year} {2020})}\BibitemShut {NoStop}%
\bibitem [{\citenamefont {Lentrodt}\ \emph {et~al.}(2020)\citenamefont {Lentrodt}, \citenamefont {Heeg}, \citenamefont {Keitel},\ and\ \citenamefont {Evers}}]{Lentrodt2020Jun}%
  \BibitemOpen
  \bibfield  {author} {\bibinfo {author} {\bibfnamefont {D.}~\bibnamefont {Lentrodt}}, \bibinfo {author} {\bibfnamefont {K.~P.}\ \bibnamefont {Heeg}}, \bibinfo {author} {\bibfnamefont {C.~H.}\ \bibnamefont {Keitel}},\ and\ \bibinfo {author} {\bibfnamefont {J.}~\bibnamefont {Evers}},\ }\bibfield  {title} {\bibinfo {title} {{Ab initio quantum models for thin-film x-ray cavity QED}},\ }\href {https://doi.org/10.1103/PhysRevResearch.2.023396} {\bibfield  {journal} {\bibinfo  {journal} {Phys. Rev. Res.}\ }\textbf {\bibinfo {volume} {2}},\ \bibinfo {pages} {023396} (\bibinfo {year} {2020})}\BibitemShut {NoStop}%
\bibitem [{\citenamefont {Dutra}\ and\ \citenamefont {Nienhuis}(2000)}]{Dutra2000Oct}%
  \BibitemOpen
  \bibfield  {author} {\bibinfo {author} {\bibfnamefont {S.~M.}\ \bibnamefont {Dutra}}\ and\ \bibinfo {author} {\bibfnamefont {G.}~\bibnamefont {Nienhuis}},\ }\bibfield  {title} {\bibinfo {title} {{Derivation of a Hamiltonian for photon decay in a}},\ }\href {https://doi.org/10.1088/1464-4266/2/5/305} {\bibfield  {journal} {\bibinfo  {journal} {J. Opt. B: Quantum Semiclassical Opt.}\ }\textbf {\bibinfo {volume} {2}},\ \bibinfo {pages} {584} (\bibinfo {year} {2000})}\BibitemShut {NoStop}%
\bibitem [{\citenamefont {Raymer}\ and\ \citenamefont {McKinstrie}(2013)}]{PhysRevA.88.043819}%
  \BibitemOpen
  \bibfield  {author} {\bibinfo {author} {\bibfnamefont {M.~G.}\ \bibnamefont {Raymer}}\ and\ \bibinfo {author} {\bibfnamefont {C.~J.}\ \bibnamefont {McKinstrie}},\ }\bibfield  {title} {\bibinfo {title} {Quantum input-output theory for optical cavities with arbitrary coupling strength: Application to two-photon wave-packet shaping},\ }\href {https://doi.org/10.1103/PhysRevA.88.043819} {\bibfield  {journal} {\bibinfo  {journal} {Phys. Rev. A}\ }\textbf {\bibinfo {volume} {88}},\ \bibinfo {pages} {043819} (\bibinfo {year} {2013})}\BibitemShut {NoStop}%
\bibitem [{\citenamefont {Khanbekyan}\ \emph {et~al.}(2005)\citenamefont {Khanbekyan}, \citenamefont {Kn\"oll}, \citenamefont {Welsch}, \citenamefont {Semenov},\ and\ \citenamefont {Vogel}}]{PhysRevA.72.053813}%
  \BibitemOpen
  \bibfield  {author} {\bibinfo {author} {\bibfnamefont {M.}~\bibnamefont {Khanbekyan}}, \bibinfo {author} {\bibfnamefont {L.}~\bibnamefont {Kn\"oll}}, \bibinfo {author} {\bibfnamefont {D.-G.}\ \bibnamefont {Welsch}}, \bibinfo {author} {\bibfnamefont {A.~A.}\ \bibnamefont {Semenov}},\ and\ \bibinfo {author} {\bibfnamefont {W.}~\bibnamefont {Vogel}},\ }\bibfield  {title} {\bibinfo {title} {Qed of lossy cavities: Operator and quantum-state input-output relations},\ }\href {https://doi.org/10.1103/PhysRevA.72.053813} {\bibfield  {journal} {\bibinfo  {journal} {Phys. Rev. A}\ }\textbf {\bibinfo {volume} {72}},\ \bibinfo {pages} {053813} (\bibinfo {year} {2005})}\BibitemShut {NoStop}%
\bibitem [{\citenamefont {Viviescas}\ and\ \citenamefont {Hackenbroich}(2003)}]{Viviescas2003Jan}%
  \BibitemOpen
  \bibfield  {author} {\bibinfo {author} {\bibfnamefont {C.}~\bibnamefont {Viviescas}}\ and\ \bibinfo {author} {\bibfnamefont {G.}~\bibnamefont {Hackenbroich}},\ }\bibfield  {title} {\bibinfo {title} {{Field quantization for open optical cavities}},\ }\href {https://doi.org/10.1103/PhysRevA.67.013805} {\bibfield  {journal} {\bibinfo  {journal} {Phys. Rev. A}\ }\textbf {\bibinfo {volume} {67}},\ \bibinfo {pages} {013805} (\bibinfo {year} {2003})}\BibitemShut {NoStop}%
\bibitem [{\citenamefont {Lednev}\ \emph {et~al.}(2024)\citenamefont {Lednev}, \citenamefont {Garc\'{\i}a-Vidal},\ and\ \citenamefont {Feist}}]{Lednev2023May}%
  \BibitemOpen
  \bibfield  {author} {\bibinfo {author} {\bibfnamefont {M.}~\bibnamefont {Lednev}}, \bibinfo {author} {\bibfnamefont {F.~J.}\ \bibnamefont {Garc\'{\i}a-Vidal}},\ and\ \bibinfo {author} {\bibfnamefont {J.}~\bibnamefont {Feist}},\ }\bibfield  {title} {\bibinfo {title} {Lindblad master equation capable of describing hybrid quantum systems in the ultrastrong coupling regime},\ }\href {https://doi.org/10.1103/PhysRevLett.132.106902} {\bibfield  {journal} {\bibinfo  {journal} {Phys. Rev. Lett.}\ }\textbf {\bibinfo {volume} {132}},\ \bibinfo {pages} {106902} (\bibinfo {year} {2024})}\BibitemShut {NoStop}%
\bibitem [{\citenamefont {Lambert}\ \emph {et~al.}(2019)\citenamefont {Lambert}, \citenamefont {Ahmed}, \citenamefont {Cirio},\ and\ \citenamefont {Nori}}]{Lambert2019Aug}%
  \BibitemOpen
  \bibfield  {author} {\bibinfo {author} {\bibfnamefont {N.}~\bibnamefont {Lambert}}, \bibinfo {author} {\bibfnamefont {S.}~\bibnamefont {Ahmed}}, \bibinfo {author} {\bibfnamefont {M.}~\bibnamefont {Cirio}},\ and\ \bibinfo {author} {\bibfnamefont {F.}~\bibnamefont {Nori}},\ }\bibfield  {title} {\bibinfo {title} {{Modelling the ultra-strongly coupled spin-boson model with unphysical modes}},\ }\href {https://doi.org/10.1038/s41467-019-11656-1} {\bibfield  {journal} {\bibinfo  {journal} {Nat. Commun.}\ }\textbf {\bibinfo {volume} {10}},\ \bibinfo {pages} {1} (\bibinfo {year} {2019})}\BibitemShut {NoStop}%
\bibitem [{\citenamefont {Dung}\ \emph {et~al.}(1998)\citenamefont {Dung}, \citenamefont {Kn{\ifmmode\ddot{o}\else\"{o}\fi}ll},\ and\ \citenamefont {Welsch}}]{Dung1998May}%
  \BibitemOpen
  \bibfield  {author} {\bibinfo {author} {\bibfnamefont {H.~T.}\ \bibnamefont {Dung}}, \bibinfo {author} {\bibfnamefont {L.}~\bibnamefont {Kn{\ifmmode\ddot{o}\else\"{o}\fi}ll}},\ and\ \bibinfo {author} {\bibfnamefont {D.-G.}\ \bibnamefont {Welsch}},\ }\bibfield  {title} {\bibinfo {title} {{Three-dimensional quantization of the electromagnetic field in dispersive and absorbing inhomogeneous dielectrics}},\ }\href {https://doi.org/10.1103/PhysRevA.57.3931} {\bibfield  {journal} {\bibinfo  {journal} {Phys. Rev. A}\ }\textbf {\bibinfo {volume} {57}},\ \bibinfo {pages} {3931} (\bibinfo {year} {1998})}\BibitemShut {NoStop}%
\bibitem [{\citenamefont {Gustin}\ \emph {et~al.}(2023)\citenamefont {Gustin}, \citenamefont {Franke},\ and\ \citenamefont {Hughes}}]{Gustin2023Jan}%
  \BibitemOpen
  \bibfield  {author} {\bibinfo {author} {\bibfnamefont {C.}~\bibnamefont {Gustin}}, \bibinfo {author} {\bibfnamefont {S.}~\bibnamefont {Franke}},\ and\ \bibinfo {author} {\bibfnamefont {S.}~\bibnamefont {Hughes}},\ }\bibfield  {title} {\bibinfo {title} {{Gauge-invariant theory of truncated quantum light-matter interactions in arbitrary media}},\ }\href {https://doi.org/10.1103/PhysRevA.107.013722} {\bibfield  {journal} {\bibinfo  {journal} {Phys. Rev. A}\ }\textbf {\bibinfo {volume} {107}},\ \bibinfo {pages} {013722} (\bibinfo {year} {2023})}\BibitemShut {NoStop}%
\bibitem [{\citenamefont {Di~Stefano}\ \emph {et~al.}(2019)\citenamefont {Di~Stefano}, \citenamefont {Settineri}, \citenamefont {Macr{\ifmmode\grave{\imath}\else\`{\i}\fi}}, \citenamefont {Garziano}, \citenamefont {Stassi}, \citenamefont {Savasta},\ and\ \citenamefont {Nori}}]{DiStefano2019Aug}%
  \BibitemOpen
  \bibfield  {author} {\bibinfo {author} {\bibfnamefont {O.}~\bibnamefont {Di~Stefano}}, \bibinfo {author} {\bibfnamefont {A.}~\bibnamefont {Settineri}}, \bibinfo {author} {\bibfnamefont {V.}~\bibnamefont {Macr{\ifmmode\grave{\imath}\else\`{\i}\fi}}}, \bibinfo {author} {\bibfnamefont {L.}~\bibnamefont {Garziano}}, \bibinfo {author} {\bibfnamefont {R.}~\bibnamefont {Stassi}}, \bibinfo {author} {\bibfnamefont {S.}~\bibnamefont {Savasta}},\ and\ \bibinfo {author} {\bibfnamefont {F.}~\bibnamefont {Nori}},\ }\bibfield  {title} {\bibinfo {title} {{Resolution of gauge ambiguities in ultrastrong-coupling cavity quantum electrodynamics}},\ }\href {https://doi.org/10.1038/s41567-019-0534-4} {\bibfield  {journal} {\bibinfo  {journal} {Nat. Phys.}\ }\textbf {\bibinfo {volume} {15}},\ \bibinfo {pages} {803} (\bibinfo {year} {2019})}\BibitemShut {NoStop}%
\bibitem [{\citenamefont {Taylor}\ \emph {et~al.}(2020)\citenamefont {Taylor}, \citenamefont {Mandal}, \citenamefont {Zhou},\ and\ \citenamefont {Huo}}]{Taylor2020Sep}%
  \BibitemOpen
  \bibfield  {author} {\bibinfo {author} {\bibfnamefont {M.~A.~D.}\ \bibnamefont {Taylor}}, \bibinfo {author} {\bibfnamefont {A.}~\bibnamefont {Mandal}}, \bibinfo {author} {\bibfnamefont {W.}~\bibnamefont {Zhou}},\ and\ \bibinfo {author} {\bibfnamefont {P.}~\bibnamefont {Huo}},\ }\bibfield  {title} {\bibinfo {title} {{Resolution of Gauge Ambiguities in Molecular Cavity Quantum Electrodynamics}},\ }\href {https://doi.org/10.1103/PhysRevLett.125.123602} {\bibfield  {journal} {\bibinfo  {journal} {Phys. Rev. Lett.}\ }\textbf {\bibinfo {volume} {125}},\ \bibinfo {pages} {123602} (\bibinfo {year} {2020})}\BibitemShut {NoStop}%
\bibitem [{\citenamefont {Settineri}\ \emph {et~al.}(2021)\citenamefont {Settineri}, \citenamefont {Di~Stefano}, \citenamefont {Zueco}, \citenamefont {Hughes}, \citenamefont {Savasta},\ and\ \citenamefont {Nori}}]{Settineri2021Apr}%
  \BibitemOpen
  \bibfield  {author} {\bibinfo {author} {\bibfnamefont {A.}~\bibnamefont {Settineri}}, \bibinfo {author} {\bibfnamefont {O.}~\bibnamefont {Di~Stefano}}, \bibinfo {author} {\bibfnamefont {D.}~\bibnamefont {Zueco}}, \bibinfo {author} {\bibfnamefont {S.}~\bibnamefont {Hughes}}, \bibinfo {author} {\bibfnamefont {S.}~\bibnamefont {Savasta}},\ and\ \bibinfo {author} {\bibfnamefont {F.}~\bibnamefont {Nori}},\ }\bibfield  {title} {\bibinfo {title} {{Gauge freedom, quantum measurements, and time-dependent interactions in cavity QED}},\ }\href {https://doi.org/10.1103/PhysRevResearch.3.023079} {\bibfield  {journal} {\bibinfo  {journal} {Phys. Rev. Res.}\ }\textbf {\bibinfo {volume} {3}},\ \bibinfo {pages} {023079} (\bibinfo {year} {2021})}\BibitemShut {NoStop}%
\bibitem [{\citenamefont {Savasta}\ \emph {et~al.}(2021)\citenamefont {Savasta}, \citenamefont {Di~Stefano}, \citenamefont {Settineri}, \citenamefont {Zueco}, \citenamefont {Hughes},\ and\ \citenamefont {Nori}}]{Savasta2021May}%
  \BibitemOpen
  \bibfield  {author} {\bibinfo {author} {\bibfnamefont {S.}~\bibnamefont {Savasta}}, \bibinfo {author} {\bibfnamefont {O.}~\bibnamefont {Di~Stefano}}, \bibinfo {author} {\bibfnamefont {A.}~\bibnamefont {Settineri}}, \bibinfo {author} {\bibfnamefont {D.}~\bibnamefont {Zueco}}, \bibinfo {author} {\bibfnamefont {S.}~\bibnamefont {Hughes}},\ and\ \bibinfo {author} {\bibfnamefont {F.}~\bibnamefont {Nori}},\ }\bibfield  {title} {\bibinfo {title} {{Gauge principle and gauge invariance in two-level systems}},\ }\href {https://doi.org/10.1103/PhysRevA.103.053703} {\bibfield  {journal} {\bibinfo  {journal} {Phys. Rev. A}\ }\textbf {\bibinfo {volume} {103}},\ \bibinfo {pages} {053703} (\bibinfo {year} {2021})}\BibitemShut {NoStop}%
\bibitem [{\citenamefont {Taylor}\ \emph {et~al.}(2022)\citenamefont {Taylor}, \citenamefont {Taylor}, \citenamefont {Mandal}, \citenamefont {Mandal}, \citenamefont {Huo},\ and\ \citenamefont {Huo}}]{Taylor2022May}%
  \BibitemOpen
  \bibfield  {author} {\bibinfo {author} {\bibfnamefont {M.~A.~D.}\ \bibnamefont {Taylor}}, \bibinfo {author} {\bibfnamefont {M.~A.~D.}\ \bibnamefont {Taylor}}, \bibinfo {author} {\bibfnamefont {A.}~\bibnamefont {Mandal}}, \bibinfo {author} {\bibfnamefont {A.}~\bibnamefont {Mandal}}, \bibinfo {author} {\bibfnamefont {P.}~\bibnamefont {Huo}},\ and\ \bibinfo {author} {\bibfnamefont {P.}~\bibnamefont {Huo}},\ }\bibfield  {title} {\bibinfo {title} {{Resolution of Gauge Ambiguities in Cavity Quantum Electrodynamics}},\ }\href {https://doi.org/10.1364/CLEO_AT.2022.JTu3A.6} {\bibfield  {journal} {\bibinfo  {journal} {Optica Publishing Group}\ ,\ \bibinfo {pages} {JTu3A.6}} (\bibinfo {year} {2022})}\BibitemShut {NoStop}%
\bibitem [{Note1()}]{Note1}%
  \BibitemOpen
  \bibinfo {note} {{Here we have neglected longitudinal field couplings~\cite {Gustin2023Jan,Knoll2000Jun} for simplicity, assuming the transverse couplings to dominate for the single-mode systems of interest, which we also verify in the Supplementary Material~\cite {SI}}}\BibitemShut {NoStop}%
\bibitem [{\citenamefont {Breuer}\ and\ \citenamefont {Petruccione}(2007)}]{Breuer2007Mar}%
  \BibitemOpen
  \bibfield  {author} {\bibinfo {author} {\bibfnamefont {H.-P.}\ \bibnamefont {Breuer}}\ and\ \bibinfo {author} {\bibfnamefont {F.}~\bibnamefont {Petruccione}},\ }\href {https://www.amazon.com/Theory-Open-Quantum-Systems/dp/0199213909} {\emph {\bibinfo {title} {{The Theory of Open Quantum Systems}}}}\ (\bibinfo  {publisher} {Oxford University Press},\ \bibinfo {address} {Oxford, England, UK},\ \bibinfo {year} {2007})\BibitemShut {NoStop}%
\bibitem [{\citenamefont {Kristensen}\ \emph {et~al.}(2020)\citenamefont {Kristensen}, \citenamefont {Herrmann}, \citenamefont {Intravaia}, \citenamefont {Busch},\ and\ \citenamefont {Busch}}]{Kristensen2020Sep}%
  \BibitemOpen
  \bibfield  {author} {\bibinfo {author} {\bibfnamefont {P.~T.}\ \bibnamefont {Kristensen}}, \bibinfo {author} {\bibfnamefont {K.}~\bibnamefont {Herrmann}}, \bibinfo {author} {\bibfnamefont {F.}~\bibnamefont {Intravaia}}, \bibinfo {author} {\bibfnamefont {K.}~\bibnamefont {Busch}},\ and\ \bibinfo {author} {\bibfnamefont {K.}~\bibnamefont {Busch}},\ }\bibfield  {title} {\bibinfo {title} {{Modeling electromagnetic resonators using quasinormal modes}},\ }\href {https://doi.org/10.1364/AOP.377940} {\bibfield  {journal} {\bibinfo  {journal} {Adv. Opt. Photonics}\ }\textbf {\bibinfo {volume} {12}},\ \bibinfo {pages} {612} (\bibinfo {year} {2020})}\BibitemShut {NoStop}%
\bibitem [{SI()}]{SI}%
  \BibitemOpen
  \href@noop {} {}\bibinfo {note} {See Supplemental Material at [URL will be inserted by publisher] for more detailed information, where we also reference papers~\cite{Chikkaraddy2016Jul,Gustin2018Jul,ren_near-field_2020,2017PRA_hybrid,bai_efficient_2013,Novotny2007Principles,jun_nonresonant_2008,liu_excitation_2009,ren_classical_2023, Franke2019May}}\BibitemShut {NoStop}%
\bibitem [{\citenamefont {Franke}\ \emph {et~al.}(2020)\citenamefont {Franke}, \citenamefont {Ren}, \citenamefont {Hughes},\ and\ \citenamefont {Richter}}]{franke_fluctuation-dissipation_2020}%
  \BibitemOpen
  \bibfield  {author} {\bibinfo {author} {\bibfnamefont {S.}~\bibnamefont {Franke}}, \bibinfo {author} {\bibfnamefont {J.}~\bibnamefont {Ren}}, \bibinfo {author} {\bibfnamefont {S.}~\bibnamefont {Hughes}},\ and\ \bibinfo {author} {\bibfnamefont {M.}~\bibnamefont {Richter}},\ }\bibfield  {title} {\bibinfo {title} {Fluctuation-dissipation theorem and fundamental photon commutation relations in lossy nanostructures using quasinormal modes},\ }\href {https://doi.org/10.1103/PhysRevResearch.2.033332} {\bibfield  {journal} {\bibinfo  {journal} {Phys. Rev. Res.}\ }\textbf {\bibinfo {volume} {2}},\ \bibinfo {pages} {033332} (\bibinfo {year} {2020})}\BibitemShut {NoStop}%
\bibitem [{\citenamefont {Blais}\ \emph {et~al.}(2021)\citenamefont {Blais}, \citenamefont {Grimsmo}, \citenamefont {Girvin},\ and\ \citenamefont {Wallraff}}]{Blais2021May}%
  \BibitemOpen
  \bibfield  {author} {\bibinfo {author} {\bibfnamefont {A.}~\bibnamefont {Blais}}, \bibinfo {author} {\bibfnamefont {A.~L.}\ \bibnamefont {Grimsmo}}, \bibinfo {author} {\bibfnamefont {S.~M.}\ \bibnamefont {Girvin}},\ and\ \bibinfo {author} {\bibfnamefont {A.}~\bibnamefont {Wallraff}},\ }\bibfield  {title} {\bibinfo {title} {{Circuit quantum electrodynamics}},\ }\href {https://doi.org/10.1103/RevModPhys.93.025005} {\bibfield  {journal} {\bibinfo  {journal} {Rev. Mod. Phys.}\ }\textbf {\bibinfo {volume} {93}},\ \bibinfo {pages} {025005} (\bibinfo {year} {2021})}\BibitemShut {NoStop}%
\bibitem [{\citenamefont {Knoll}\ \emph {et~al.}(2000)\citenamefont {Knoll}, \citenamefont {Scheel},\ and\ \citenamefont {Welsch}}]{Knoll2000Jun}%
  \BibitemOpen
  \bibfield  {author} {\bibinfo {author} {\bibfnamefont {L.}~\bibnamefont {Knoll}}, \bibinfo {author} {\bibfnamefont {S.}~\bibnamefont {Scheel}},\ and\ \bibinfo {author} {\bibfnamefont {D.-G.}\ \bibnamefont {Welsch}},\ }\bibfield  {title} {\bibinfo {title} {{QED in dispersing and absorbing media}},\ }\bibfield  {journal} {\bibinfo  {journal} {arXiv}\ }\href {https://doi.org/10.48550/arXiv.quant-ph/0006121} {10.48550/arXiv.quant-ph/0006121} (\bibinfo {year} {2000}),\ \Eprint {https://arxiv.org/abs/quant-ph/0006121} {quant-ph/0006121} \BibitemShut {NoStop}%
\bibitem [{\citenamefont {Chikkaraddy}\ \emph {et~al.}(2016)\citenamefont {Chikkaraddy}, \citenamefont {de~Nijs}, \citenamefont {Benz}, \citenamefont {Barrow}, \citenamefont {Scherman}, \citenamefont {Rosta}, \citenamefont {Demetriadou}, \citenamefont {Fox}, \citenamefont {Hess},\ and\ \citenamefont {Baumberg}}]{Chikkaraddy2016Jul}%
  \BibitemOpen
  \bibfield  {author} {\bibinfo {author} {\bibfnamefont {R.}~\bibnamefont {Chikkaraddy}}, \bibinfo {author} {\bibfnamefont {B.}~\bibnamefont {de~Nijs}}, \bibinfo {author} {\bibfnamefont {F.}~\bibnamefont {Benz}}, \bibinfo {author} {\bibfnamefont {S.~J.}\ \bibnamefont {Barrow}}, \bibinfo {author} {\bibfnamefont {O.~A.}\ \bibnamefont {Scherman}}, \bibinfo {author} {\bibfnamefont {E.}~\bibnamefont {Rosta}}, \bibinfo {author} {\bibfnamefont {A.}~\bibnamefont {Demetriadou}}, \bibinfo {author} {\bibfnamefont {P.}~\bibnamefont {Fox}}, \bibinfo {author} {\bibfnamefont {O.}~\bibnamefont {Hess}},\ and\ \bibinfo {author} {\bibfnamefont {J.~J.}\ \bibnamefont {Baumberg}},\ }\bibfield  {title} {\bibinfo {title} {{Single-molecule strong coupling at room temperature in plasmonic nanocavities}},\ }\href {https://doi.org/10.1038/nature17974} {\bibfield  {journal} {\bibinfo  {journal} {Nature}\ }\textbf {\bibinfo {volume} {535}},\ \bibinfo {pages} {127} (\bibinfo {year} {2016})}\BibitemShut {NoStop}%
\bibitem [{\citenamefont {Gustin}\ and\ \citenamefont {Hughes}(2018)}]{Gustin2018Jul}%
  \BibitemOpen
  \bibfield  {author} {\bibinfo {author} {\bibfnamefont {C.}~\bibnamefont {Gustin}}\ and\ \bibinfo {author} {\bibfnamefont {S.}~\bibnamefont {Hughes}},\ }\bibfield  {title} {\bibinfo {title} {{Pulsed excitation dynamics in quantum-dot--cavity systems: Limits to optimizing the fidelity of on-demand single-photon sources}},\ }\href {https://doi.org/10.1103/PhysRevB.98.045309} {\bibfield  {journal} {\bibinfo  {journal} {Phys. Rev. B}\ }\textbf {\bibinfo {volume} {98}},\ \bibinfo {pages} {045309} (\bibinfo {year} {2018})}\BibitemShut {NoStop}%
\bibitem [{\citenamefont {Ren}\ \emph {et~al.}(2020)\citenamefont {Ren}, \citenamefont {Franke}, \citenamefont {Knorr}, \citenamefont {Richter},\ and\ \citenamefont {Hughes}}]{ren_near-field_2020}%
  \BibitemOpen
  \bibfield  {author} {\bibinfo {author} {\bibfnamefont {J.}~\bibnamefont {Ren}}, \bibinfo {author} {\bibfnamefont {S.}~\bibnamefont {Franke}}, \bibinfo {author} {\bibfnamefont {A.}~\bibnamefont {Knorr}}, \bibinfo {author} {\bibfnamefont {M.}~\bibnamefont {Richter}},\ and\ \bibinfo {author} {\bibfnamefont {S.}~\bibnamefont {Hughes}},\ }\bibfield  {title} {\bibinfo {title} {Near-field to far-field transformations of optical quasinormal modes and efficient calculation of quantized quasinormal modes for open cavities and plasmonic resonators},\ }\href {https://doi.org/10.1103/PhysRevB.101.205402} {\bibfield  {journal} {\bibinfo  {journal} {Phys. Rev. B}\ }\textbf {\bibinfo {volume} {101}},\ \bibinfo {pages} {205402} (\bibinfo {year} {2020})}\BibitemShut {NoStop}%
\bibitem [{\citenamefont {Kamandar~Dezfouli}\ \emph {et~al.}(2017)\citenamefont {Kamandar~Dezfouli}, \citenamefont {Gordon},\ and\ \citenamefont {Hughes}}]{2017PRA_hybrid}%
  \BibitemOpen
  \bibfield  {author} {\bibinfo {author} {\bibfnamefont {M.}~\bibnamefont {Kamandar~Dezfouli}}, \bibinfo {author} {\bibfnamefont {R.}~\bibnamefont {Gordon}},\ and\ \bibinfo {author} {\bibfnamefont {S.}~\bibnamefont {Hughes}},\ }\bibfield  {title} {\bibinfo {title} {Modal theory of modified spontaneous emission of a quantum emitter in a hybrid plasmonic photonic-crystal cavity system},\ }\href {https://doi.org/10.1103/PhysRevA.95.013846} {\bibfield  {journal} {\bibinfo  {journal} {Phys. Rev. A}\ }\textbf {\bibinfo {volume} {95}},\ \bibinfo {pages} {013846} (\bibinfo {year} {2017})}\BibitemShut {NoStop}%
\bibitem [{\citenamefont {Bai}\ \emph {et~al.}(2013)\citenamefont {Bai}, \citenamefont {Perrin}, \citenamefont {Sauvan}, \citenamefont {Hugonin},\ and\ \citenamefont {Lalanne}}]{bai_efficient_2013}%
  \BibitemOpen
  \bibfield  {author} {\bibinfo {author} {\bibfnamefont {Q.}~\bibnamefont {Bai}}, \bibinfo {author} {\bibfnamefont {M.}~\bibnamefont {Perrin}}, \bibinfo {author} {\bibfnamefont {C.}~\bibnamefont {Sauvan}}, \bibinfo {author} {\bibfnamefont {J.-P.}\ \bibnamefont {Hugonin}},\ and\ \bibinfo {author} {\bibfnamefont {P.}~\bibnamefont {Lalanne}},\ }\bibfield  {title} {\bibinfo {title} {Efficient and intuitive method for the analysis of light scattering by a resonant nanostructure},\ }\href {https://doi.org/10.1364/oe.21.027371} {\bibfield  {journal} {\bibinfo  {journal} {Optics Express}\ }\textbf {\bibinfo {volume} {21}},\ \bibinfo {pages} {27371} (\bibinfo {year} {2013})}\BibitemShut {NoStop}%
\bibitem [{\citenamefont {Novotny}\ and\ \citenamefont {Hecht}(2006)}]{Novotny2007Principles}%
  \BibitemOpen
  \bibfield  {author} {\bibinfo {author} {\bibfnamefont {L.}~\bibnamefont {Novotny}}\ and\ \bibinfo {author} {\bibfnamefont {B.}~\bibnamefont {Hecht}},\ }\href@noop {} {\emph {\bibinfo {title} {Principles of Nano{-}Optics}}}\ (\bibinfo  {publisher} {Cambridge University Press, New York},\ \bibinfo {year} {2006})\BibitemShut {NoStop}%
\bibitem [{\citenamefont {Jun}\ \emph {et~al.}(2008)\citenamefont {Jun}, \citenamefont {Kekatpure}, \citenamefont {White},\ and\ \citenamefont {Brongersma}}]{jun_nonresonant_2008}%
  \BibitemOpen
  \bibfield  {author} {\bibinfo {author} {\bibfnamefont {Y.~C.}\ \bibnamefont {Jun}}, \bibinfo {author} {\bibfnamefont {R.~D.}\ \bibnamefont {Kekatpure}}, \bibinfo {author} {\bibfnamefont {J.~S.}\ \bibnamefont {White}},\ and\ \bibinfo {author} {\bibfnamefont {M.~L.}\ \bibnamefont {Brongersma}},\ }\bibfield  {title} {\bibinfo {title} {Nonresonant enhancement of spontaneous emission in metal-dielectric-metal plasmon waveguide structures},\ }\href {https://doi.org/10.1103/PhysRevB.78.153111} {\bibfield  {journal} {\bibinfo  {journal} {Physical Review B}\ }\textbf {\bibinfo {volume} {78}},\ \bibinfo {pages} {153111} (\bibinfo {year} {2008})}\BibitemShut {NoStop}%
\bibitem [{\citenamefont {Liu}\ \emph {et~al.}(2009)\citenamefont {Liu}, \citenamefont {Lee}, \citenamefont {Gray}, \citenamefont {Guyot-Sionnest},\ and\ \citenamefont {Pelton}}]{liu_excitation_2009}%
  \BibitemOpen
  \bibfield  {author} {\bibinfo {author} {\bibfnamefont {M.}~\bibnamefont {Liu}}, \bibinfo {author} {\bibfnamefont {T.-W.}\ \bibnamefont {Lee}}, \bibinfo {author} {\bibfnamefont {S.~K.}\ \bibnamefont {Gray}}, \bibinfo {author} {\bibfnamefont {P.}~\bibnamefont {Guyot-Sionnest}},\ and\ \bibinfo {author} {\bibfnamefont {M.}~\bibnamefont {Pelton}},\ }\bibfield  {title} {\bibinfo {title} {Excitation of dark plasmons in metal nanoparticles by a localized emitter},\ }\href {https://doi.org/10.1103/PhysRevLett.102.107401} {\bibfield  {journal} {\bibinfo  {journal} {Physical Review Letters}\ }\textbf {\bibinfo {volume} {102}},\ \bibinfo {pages} {107401} (\bibinfo {year} {2009})}\BibitemShut {NoStop}%
\bibitem [{\citenamefont {Ren}\ \emph {et~al.}(2024)\citenamefont {Ren}, \citenamefont {Franke}, \citenamefont {VanDrunen},\ and\ \citenamefont {Hughes}}]{ren_classical_2023}%
  \BibitemOpen
  \bibfield  {author} {\bibinfo {author} {\bibfnamefont {J.}~\bibnamefont {Ren}}, \bibinfo {author} {\bibfnamefont {S.}~\bibnamefont {Franke}}, \bibinfo {author} {\bibfnamefont {B.}~\bibnamefont {VanDrunen}},\ and\ \bibinfo {author} {\bibfnamefont {S.}~\bibnamefont {Hughes}},\ }\bibfield  {title} {\bibinfo {title} {Classical {Purcell} factors and spontaneous emission decay rates in a linear gain medium},\ }\href {https://doi.org/10.1103/PhysRevA.109.013513} {\bibfield  {journal} {\bibinfo  {journal} {Phys. Rev. A}\ }\textbf {\bibinfo {volume} {109}},\ \bibinfo {pages} {013513} (\bibinfo {year} {2024})}\BibitemShut {NoStop}%
\bibitem [{\citenamefont {Franke}\ \emph {et~al.}(2019)\citenamefont {Franke}, \citenamefont {Hughes}, \citenamefont {Dezfouli}, \citenamefont {Kristensen}, \citenamefont {Busch}, \citenamefont {Knorr},\ and\ \citenamefont {Richter}}]{Franke2019May}%
  \BibitemOpen
  \bibfield  {author} {\bibinfo {author} {\bibfnamefont {S.}~\bibnamefont {Franke}}, \bibinfo {author} {\bibfnamefont {S.}~\bibnamefont {Hughes}}, \bibinfo {author} {\bibfnamefont {M.~K.}\ \bibnamefont {Dezfouli}}, \bibinfo {author} {\bibfnamefont {P.~T.}\ \bibnamefont {Kristensen}}, \bibinfo {author} {\bibfnamefont {K.}~\bibnamefont {Busch}}, \bibinfo {author} {\bibfnamefont {A.}~\bibnamefont {Knorr}},\ and\ \bibinfo {author} {\bibfnamefont {M.}~\bibnamefont {Richter}},\ }\bibfield  {title} {\bibinfo {title} {{Quantization of Quasinormal Modes for Open Cavities and Plasmonic Cavity Quantum Electrodynamics}},\ }\href {https://doi.org/10.1103/PhysRevLett.122.213901} {\bibfield  {journal} {\bibinfo  {journal} {Phys. Rev. Lett.}\ }\textbf {\bibinfo {volume} {122}},\ \bibinfo {pages} {213901} (\bibinfo {year} {2019})}\BibitemShut {NoStop}%
\end{thebibliography}%
\end{document}